\pdfoutput=1
\documentclass[reqno, 11pt]{amsart}

\usepackage[english]{babel}
\usepackage[T1]{fontenc}
\usepackage[utf8]{inputenc}

\usepackage{amsfonts,amssymb,amsbsy,amsmath,amsthm,enumerate}
\usepackage{hyperref}
\usepackage{eucal}
\usepackage{color}
\usepackage{mathrsfs}
\usepackage[small,nohug,
]{diagrams}
\diagramstyle[labelstyle=\scriptstyle]
\newarrow{Corresponds}<--->

\usepackage{lxfonts}
\theoremstyle{plain}
\newtheorem{theorem}{Theorem}[section]
\theoremstyle{plain}
\newtheorem{corollary}{Corollary}[section]
\theoremstyle{plain}
\newtheorem{proposition}{Proposition}[section]
\theoremstyle{plain}
\newtheorem{lemma}{Lemma}[section]
\theoremstyle{definition}
\newtheorem{definition}{Definition}[section]
\theoremstyle{definition}

\theoremstyle{definition}

\theoremstyle{definition}

\theoremstyle{definition}

\theoremstyle{definition}
\newtheorem{remark}{Remark}[section]

\numberwithin{equation}{section}
\numberwithin{figure}{section}
\numberwithin{table}{section}

\newcommand{\R}{\mathbb{R}}
\newcommand{\N}{\mathbb{N}}
\newcommand{\C}{\mathbb{C}}                           %NUMBER SETS
\newcommand{\Q}{\mathbb{Q}}
\newcommand{\Z}{\mathbb{Z}}

\newcommand{\s}[1]{\CMcal{#1}}
\newcommand{\bb}[1]{\mathscr{#1}}
\newcommand{\rr}[1]{\mathfrak{#1}}
\newcommand{\n}[1]{\mathbb{#1}}

\newcommand{\expo}[1]{\,\mathrm{e}^{#1}\,}                 %SPECIAL FUNCTION
\newcommand{\dd}{\,\mathrm{d}}
\newcommand{\ii}{\,\mathrm{i}\,}

\newcommand{\virg}[1]{\lq\lq#1\rq\rq}                %TIPOGRAPHIC
\newcommand{\ie}{\textsl{i.\,e.\,}}
\newcommand{\eg}{\textsl{e.\,g.\,}}
\newcommand{\cf}{\textsl{cf}.\,}

\newcommand{\lp}{\left(}
\newcommand{\rp}{\right)}

\title[Spectral Theory of the Thermal Hamiltonian: 1D Case]{Spectral Theory of the Thermal Hamiltonian: 1D Case}

\author[G. De~Nittis]{Giuseppe De Nittis}

\address[G. De~Nittis]{Facultad de Matem\'aticas \& Instituto de F\'{\i}sica,
  Pontificia Universidad Cat\'olica de Chile,
  Santiago, Chile.}
\email{gidenittis@mat.uc.cl}

\author[V. Lenz]{Vicente Lenz}

\address[V. Lenz]{Departamento de Matem\'aticas, Facultad de Ciencias,  Universidad de Chile, Santiago, Chile}
\email{vicente.lenz@ug.uchile.cl}


\begin{document}

\begin{abstract}
In 1964 J. M. Luttinger introduced a model for the quantum thermal transport. In this paper we study the spectral theory of the Hamiltonian operator associated to  the Luttinger's model, with a special focus at the one-dimensional case. It is shown that the (so called) thermal Hamiltonian has a one-parameter family of self-adjoint extensions and the 
spectrum, the time-propagator group and the Green function are explicitly computed. Moreover, the scattering by convolution-type potentials is analyzed. 
Finally, also the associated classical problem is completely solved, thus providing a comparison between classical and quantum behavior. This article aims to be a first contribution in the construction of a complete theory for the  thermal Hamiltonian.
\end{abstract}

\maketitle

\vspace{-5mm}
\tableofcontents

%--------------------%
%--------------------%
\section{Introduction}\label{sect:intro}
The aim of this introductory section is twice: First of all, we will provide the physical background that motivates the study of  the Thermal Hamiltonian; Secondly, we will present the mathematical problems and the main results achieved in this work.

%-----%
\subsection{Physical motivations}
The motion of an electron inside the matter, and 
subjected to a static magnetic field ${B}$,  is described by the (one-particle) Hamiltonian
\begin{equation}\label{eq:intro_01}
H(A,V)\;:=\;K(A)\;+\;V
\end{equation}
where
\begin{equation}\label{eq:intro_022202}
K(A)\;:=\;\frac{1}{2m}\left(p\;-\;\frac{e}{c}A\right)^2\;.
\end{equation}
The parameters $m$ and $e$ describe the \emph{mass} and the \emph{charge} of the electron, respectively. The constant $c$ is the (in vacum) \emph{speed of the light}.
The static (effective) potential $V$  takes care of the interaction of the electron with the atomic structure of the matter and causes only \emph{elastic scattering}. The magnetic field enters in the \emph{kinetic} term  $K(A)$ through its \emph{vector potential} $A$ according to the equation $B=\nabla\times A$. In  Quantum Mechanics the Hamiltonian $H(A,V)$ is interpreted as a differential operator acting on the Hilbert space $L^2(\R^d)$, where the differential part is provided by the momentum operator $p:=-\hslash \ii\nabla$, $\hslash$ being the \emph{Planck constant}.
The potentials $V=V(x)$ and  $A=A(x)$ are functions of the position operator $x=(x_1,\ldots,x_d)$, and act as multiplication operators.

\medskip

The \emph{transport phenomena} in the matter are analyzed by studing the
response of the system to an external \emph{perturbation}  $F=F(x)$ \cite{luttinger-68,luttinger-64}. In the stationary regime, that is when all the transient effects due to the switching-on of the perturbation are suppressed, the system reacts  by generating a
(stationary) \emph{drift current}. The latter can be computed (at least in the linear response regime, see \eg \cite{denittis-lein-book}) starting from the full dynamics generated by the \emph{perturbed} Hamiltonian
\begin{equation}\label{eq:intro_02}
H(A,F,V)\;:=\;H_0(A,F)\;+\;V\;.
\end{equation}
In \eqref{eq:intro_02}
  the \virg{free} Hamiltonian
\begin{equation}\label{eq:intro_03}
H_0(A,F)\;:=\;K(A)\;+\;F
\end{equation}
describes the motion of an electron that moves in the empty space
under the influence of the (external) fields  generated by $A$ and $F$. The potential $V$ in  \eqref{eq:intro_02} describes the  interaction with the  matter which generates elastic scattering of the particle. Once the \virg{free} dynamics generated by $H_0(A,F)$ is known, one can study the
influence of the matter by means of the \emph{scattering theory} \cite{reed-simon-III,yafaev-92,kato-95} for the pair of operators $H_0(A,F)$ and $H(A,F,V)$.

\medskip

The best studied case concerns the response of the system to the perturbation induced by a uniform  electric field $E=(E_1,\ldots,E_d)$. In this case the perturbation  is described  by the electrostatic potential 
$$
F_E(x)\;:=\;-eE\cdot x\;=\;-e(E_1x_1+\ldots+E_dx_d)
$$ 
and the associate perturbed Hamiltonian takes the form
\begin{equation}\label{eq:intro_02_bis}
H(A,F_E,V)\;:=\;H_{\rm Stark}(A)\;+\;V\;
\end{equation}
where the \virg{free} part is given by
\begin{equation}\label{eq:intro_04}
H_{\rm Stark}(A)\;:=\;K(A)\;-\;eE\cdot x\;
\end{equation}
according to \eqref{eq:intro_03}.
The operator $H_{\rm Stark}(A)$ is known as \emph{(magnetic) Stark Hamiltonian}. The non-magnetic case $H_{\rm Stark}(A=0)=\frac{p^2}{2m}-eE\cdot x$ has been extensively studied since the dawn of the Quantum Mechanics. Among the vast literature, we will refer to \cite{avron-herbst-77}
for a concise and rigorous presentation of the spectral theory of  $H_{\rm Stark}(0)$
and the related scattering theory when the  background potential $V$ is taken in consideration. The spectral theory of $H_{\rm Stark}(A)$ in presence of a uniform magnetic field
is discussed  in \cite{dimassi-petkov-10,assel-dimassi-fernandez-14}, among others.

\medskip

In order to study the  \emph{thermal transport} in the matter, Luttinger proposed a model which allows  a \virg{mechanical} derivation of the thermal coefficients \cite{luttinger-64}.
Such a model has been then applied and generalized successfully by other authors like in \cite{smrcka-streda-77,vafek-melikyan-tesanovic-01}. The essential point of the Luttinger's model is that the effect of the thermal gradient in the matter is replaced by a \virg{fictitious} gravitational field, which can be easily described  by a perturbation of the Hamiltonian in the spirit of \eqref{eq:intro_02} and \eqref{eq:intro_03}. More precisely, one assumes that the particle is
subject to a force which has the direction of the thermal gradient $\nabla \vartheta$ (where $\vartheta$ is the distribution of temperature) and which is proportional to the local content of energy divided by $c^2$ (in view of the mass-energy equivalence). The latter is given by the Hamiltonian \eqref{eq:intro_01} itself. Such a \emph{thermal-gravitational field} is given by the potential
\begin{equation}
\begin{aligned}
F_T\;:&=\;\frac{1}{2}\left[\left(\frac{\nabla \vartheta}{c^2}\cdot x\right)\; H(A,V)\;+\;H(A,V)\;\left(\frac{\nabla \vartheta}{c^2}\cdot x\right) \right]\\
&=\frac{\nabla \vartheta}{c^2}\cdot\frac{1}{2}\big\{H(A,V),x\big\}
\end{aligned}
\end{equation}
where the anti-commutator $\{\;,\;\}$ between $H(A,V)$ and $x$ is needed to make $F_T$ \emph{formally} self-adjoint (\ie symmetric). The total perturbed Hamiltonian $H(A,F_T,V)$, computed according to \eqref{eq:intro_02}, can be written as
\begin{equation}\label{eq:intro_05}
H(A,F_T,V)\;=\;H_T(A)\;+\; W(V)
\end{equation}
where the \virg{free} part, called \emph{(magnetic) thermal Hamiltonian}\footnote{An equivalently appropriate name for $H_T(A)$  could be (magnetic) \emph{Luttinger Hamiltonian}.
}, is given by
\begin{equation}\label{eq:intro_06}
H_T(A)\;:=\;K(A)\;+\; \frac{\nabla \vartheta}{c^2}\cdot\frac{1}{2}\big\{K(A),x\big\}
\end{equation}
and the effective \emph{gravitational-matter potential} reads
\begin{equation}\label{eq:intro_07}
W(V)\;:=\;\left(1+ \frac{\nabla \vartheta}{c^2}\cdot x\right)\; V\;.
\end{equation}

\medskip

The thermal Hamiltonian $H_T(A)$  is the analog of the Stark Hamiltonian  when the system
is perturbed by the gravitational-thermal field instead of the electric field. For this reason, it seems natural to look for the extension of the results valid for the Stark Hamiltonian (\eg \cite{avron-herbst-77,dimassi-petkov-10,assel-dimassi-fernandez-14}) to the case of the thermal Hamiltonian. This consists of two consecutive problems: (i) the analysis of the spectral theory of the \virg{free} operator $H_T(A)$; (ii) the study of the scattering theory for the pair  $H_T(A)$ and $H(A,F_T,V)$.
Both othese problems seem not to have been studied yet in the literature, at least to the best of our knowledge. For this reason we devote this work at the analysis  of the questions  (i) and (ii) above, in the one-dimensional case. The multi-dimensional case will be  treated in a future work.

%-----%
\subsection{Position of the  spectral problem}
\label{sec:intro_gen}
In order to formulate the problems sketched above in a rigorous mathematical setting we will make some simplifications. The most relevant concerns the absence of the magnetic field. From here on, unless otherwise indicated, we will fix $A=0$. It is worth mentioning that this is not a major restriction as long as one is interested only the one-dimensional regime. Indeed, in one spatial dimension the magnetic field is a pure gauge and can be removed with a unitary transformation\footnote{This fact can be interpreted as a consequence of the \emph{Stone-von Neumann theorem} (see \eg \cite{rosenberg-04}). Indeed, in one spatial dimension the pair $x$, $\pi_f:=p+f(x)$ necessarily meets the canonical commutation rule and so it is unitarily equivalent to the canonical pair $x,p$.}. 

\medskip

As usual in mathematics, we will normalize all the physical units: $2m=\hslash=c=e=1$. Moreover, we
will denote with $\lambda:=|\nabla \vartheta|>0$  the strength of the thermal gradient and with $\gamma:=\lambda^{-1}\nabla \vartheta \in\n{S}^{d-1}$ its direction. With these simplifications the thermal Hamiltonian reads
\begin{equation}\label{eq:intro_08}
H_T\;\equiv\;H_T(\lambda,\gamma)\;:=\;p^2\;+\; \frac{\lambda}{2}\big\{p^2,\gamma\cdot x\big\}\;.
\end{equation}
The expression \eqref{eq:intro_08} is  formal without the specification of the domain of definition of $H_T$. However, $H_T$ is evidently well defined on the space of the compactly supported smooth function $\s{C}^\infty_{\rm c}(\R^d)$ or on the Schwartz space $\s{S}(\R^d)$. On  these dense domains the operator \eqref{eq:intro_08} acts as
\begin{equation}\label{eq:intro_09}
\big(H_T\psi\big)(x)\;:=\;-\left(1+\lambda\; \gamma\cdot x\right)(\Delta\psi)(x)\;-\; \lambda\;(\gamma\cdot\nabla\psi)(x)\end{equation}
where $\Delta:=\sum_{j=1}^d{\partial_{x_j}^2}$ denotes the Laplacian 
and $\gamma\cdot\nabla:=\sum_{j=1}^d\gamma_j{\partial_{x_j}}$.
We can  simplify the last expression with the help of two unitary transformations of the Hilbert space $L^2(\R^d)$. The first one is the rotation
\begin{equation}\label{eq:intro_rot}
(R_\gamma\psi)(x)\;:=\;\psi\big(O_\gamma^{-1} x\big)\;,\qquad\quad \psi\in L^2(\R^d)
\end{equation}
where the orthogonal matrix $O_\gamma$ meets the condition $O_\gamma \gamma=(1,0,\ldots,0)$.
A short computation shows that
$$
\big(R_\gamma H_T R_\gamma^*\psi\big)(x)\;=\;-\left(1+\lambda\; x_1\right)(\Delta\psi)(x)\;-\; {\lambda}\;\frac{\partial\psi}{\partial x_1}(x)\;,
$$
where $x_1$ denotes the first component of the position vector $x=(x_1,x_\bot)\in\R^d$
and $x_\bot:=(x_2,\ldots,x_d)\in\R^{d-1}$ is its orthogonal complement. Evidently, the rotation $O_\gamma$ has the role of aligning the thermal-gravitational field along the $x_1$-axis\footnote{\label{Note_01}Clearly, in dimension $d=1$ the thermal-gravitational field is trivially aligned with the only spatiual axis and therefore $R_\gamma$ reduces to the identity.}.
The second transformation is the translation
\begin{equation}\label{eq:intro_transl}
(S_\lambda\psi)(x_1,x_\bot)\;:=\;\psi\left(x_1-\frac{1}{\lambda},x_\bot \right)\;,\qquad\quad \psi\in L^2(\R^d)
\end{equation}
and a direct calculation provides
$$
\big(S_\lambda R_\gamma H_T R_\gamma^*S_\lambda^*\psi\big)(x)\;=\;\lambda\left(-x_1(\Delta\psi)(x)\;-\; \;\frac{\partial\psi}{\partial x_1}(x)\right)\;.
$$
The operator on the  brackets
\begin{equation}\label{eq:intro_10}
(T\psi)(x)\;:=\;-x_1(\Delta\psi)(x)\;-\; \frac{\partial\psi}{\partial x_1}(x)
\end{equation}
agrees with the formal anti-commutator
\begin{equation}\label{eq:intro_11}
T\;\equiv\;\frac{1}{2}\big\{p^2,x_1\big\}\;:=\;\frac{1}{2}\big(p^2 x_1\;+\,x_1 p^2\big)
\end{equation}
when evaluated on sufficiently regular functions like $\psi\in \s{S}(\R^d)$. With a slight abuse of notation, we will often use the representation \eqref{eq:intro_11} for the operator $T$, instead of the more precise definition
\eqref{eq:intro_10}.

\medskip

The unitary equivalence between $H_T$ and $T$ implies that the spectral theory of the  thermal Hamiltonian $H_T$ can be completely recovered from the spectral theory of the operator $T$. For this reason, one is led to the problem  of determining if the operator $T$, initially defined by \eqref{eq:intro_10} on the dense domain $\s{S}(\R^d)$, admits  self-adjoint extensions and, in that case, to compute the related spectra. 

\medskip

For technical reasons, it results easier to face the equivalent problems   in the Fourier space. Let $\bb{F}:L^2(\R^d)\to L^2(\R^d)$ be the
Fourier transform defined  (just to fix the convention) by
$$
(\bb{F}\psi)(k)\;:=\;\frac{1}{(2\pi)^{\frac{d}{2}}}\int_{\R^d}\dd x\; \expo{-\ii k\cdot x}\; \psi(x)
$$
on the dense subspace $\psi\in L^1(\R^d)\cap L^2(\R^d)$. Let $\Pi:=\bb{F}T\bb{F}^*$
be the Fourier transformed version of the operator \eqref{eq:intro_10}. A direct computation shows that for $\psi\in \s{S}(\R^d)$
\begin{equation}\label{eq:intro_12}
\big(\Pi\psi\big)(x)\;:=\;\ii\left[x_1\psi(x)\;+\;x^2\frac{\partial\psi}{\partial x_1}(x)\right]\;,\qquad\quad 
\end{equation}
where $x^2:=\sum_{j=1}^d x_j^2$. The operator defined by \eqref{eq:intro_12}
agrees with the formal expression
\begin{equation}\label{eq:intro_13}
\Pi\;\equiv\;-\frac{1}{2}\big\{x^2,p_1\big\}\;:=\;-\frac{1}{2}\big(x^2 p_1\;+\,p_1 x^2\big)
\end{equation}
on sufficiently regular functions\footnote{Formula \eqref{eq:intro_13} can be formally derived from \eqref{eq:intro_11} by using the well known transformations of the canonical operators $\bb{F}p_j\bb{F}^*=x_j$ and $\bb{F}x_j\bb{F}^*=-p_j$ for all $j=1,\ldots,d$.}.

\medskip

The representation \eqref{eq:intro_13} is quite intriguing if one compares the operator $\Pi$ with the typical generator of $C_0$-groups associated to $\s{C}^\infty$-flows \cite[Chapter 4]{amrein-boutet-georgescu-96}.
At first glance, it would seem that the general theory of $C_0$-groups  applies to $\Pi$. However, a closer inspection to  the $\R$-flow associated to  $\Pi$ shows that this is not the case in general (see Section \ref{sect:1D-case_up} for more details). Therefore,
 the question of the self-adjointness of  $\Pi$ needs to be investigated with other tools.

\medskip

The first fundamental question is whether the operator $\Pi$, initially defined by \eqref{eq:intro_12} on $\s{S}(\R^d)$, admits self-adjoint extensions or not. This is fortunately true and easily demonstrable. Indeed, it is straightforward to check that  $\Pi$, as defined by
\eqref{eq:intro_12}, is \emph{symmetric} (hence \emph{closable}) on $ \s{S}(\R^d)$, \ie
$$
\langle\Pi\psi,\varphi\rangle\;=\;\langle\psi,\Pi\varphi\rangle\;,
\qquad\quad \forall\;\psi,\varphi\in \s{S}(\R^d)\;.
$$
This observation allows us to identify $\Pi$ with its closure (still denoted with the same symbol) defined on the domain
\begin{equation}\label{eq:intro_15}
\s{D}_{0}\;:=\;\overline{\s{S}(\R^d)}^{\;||\; ||_{\Pi}}
\end{equation}
obtained by the closure of $\s{S}(\R^d)$ with respect to the graph-norm
$$
||\psi||^2_{\Pi}\;:=\;||\psi||^2\;+\;|| \Pi\psi||^2\;.
$$
The existence of self-adjoint extensions of $\Pi$ is  justified by the von Neumann's criterion \cite[Theorem X.3]{reed-simon-II}. 
Let $\Upsilon$ be the  anti-unitary operator  on $L^2(\R^d)$ defined by $(\Upsilon\psi)(x)=\overline{\psi}(-x)$. The 
domains $\s{C}^\infty_{\rm c}(\R^d)$ or $\s{S}(\R^d)$ are left unchanged by $\Upsilon$ and a direct check shows that $\Upsilon\Pi=\Pi\Upsilon$ on these domains. This is sufficient to claim that:
\begin{proposition}\label{prop:int_01}
The closed symmetric operator $\Pi$ with domain $\s{D}_{0}$
admits
self-adjoint extensions.
\end{proposition}

\medskip

Proposition \ref{prop:int_01}  allows  a precise definition of the family of thermal Hamiltonians.
\begin{definition}[Thermal Hamiltonian]\label{def:intro_01}
Let $\Pi_\theta$ be a given self-adjoint extension of the operator $\Pi$ with domain $\s{D}(\Pi_\theta)\supset \s{D}_{0}$. Let $\bb{F}(\lambda,\gamma):=\bb{F}S_\lambda R_\gamma$ be the unitary operator given by the product of the Fourier transform $\bb{F}$, the translation $S_\lambda$ defined by \eqref{eq:intro_transl} and the rotation $R_\gamma$ defined by \eqref{eq:intro_rot}. Then, the associated thermal Hamiltonian is the self-adjoint operator
$$
H_{T,\theta}(\lambda,\gamma)\;:=\;\lambda\;\bb{F}(\lambda,\gamma)^*\;\Pi_\theta\;\bb{F}(\lambda,\gamma)\;,\qquad\quad \lambda>0,\;\;\gamma\in\n{S}^{d-1}
$$
defined on the domain $\s{D}(H_{T,\theta}):=\bb{F}(\lambda,\gamma) ^*[\s{D}(\Pi_\theta)]$.
\end{definition}
\medskip

 Definition \eqref{def:intro_01} reduces the question of the spectral theory of the thermal Hamiltonian to the analysis of the self-adjoint realizations of the operator $\Pi$.
This is usually done by studying the deficiency subspaces
$$
\s{K}_\pm\;:=\;{\rm Ker}(\ii\mp \Pi^*)\;.
$$
The existence of the conjugation $\Upsilon$ for $\Pi$ implies the equality of the \emph{deficiency indices}  $n_\pm:=\rm{dim}(\s{K}_\pm)$ \cite[Theorem X.3]{reed-simon-II}
which in turn ensures the existence of self-adjoint extensions. In order to build the spaces
$\s{K}_\pm$ and to compute $n_\pm$,
one needs to solve the equations $\Pi^*\psi=\pm\ii\psi$ which, in view of \eqref{eq:intro_12},  is equivalent of finding the \emph{weak solutions} \cite[Section V.4]{reed-simon-I} to the differential equations
$$
(x_1^2+x_\bot^2)\frac{\partial\psi}{\partial x_1}(x_1,x_\bot)\;+\;(x_1\mp1)\psi(x_1,x_\bot)\;=\;0\;\quad \psi\in L^2(\R^d)\cap\s{S}'(\R^d)
$$
where $\s{S}'(\R^d)$ is the space of {tempered distributions}\footnote{Similarly, one can consider weak solutions in $L^2(\R^d)\cap\s{D}'(\R^d)$ where $\s{D}'(\R^d)\supset \s{S}'(\R^d)$ is the space of distributions.}. This problem will be solved for the one-dimensional case in Section \ref{sect:self_ext1D-case}.

%-----%
\subsection{Overview on the one-dimensional case}
In Section \ref{sect:self_ext1D-case},   it is shown that  the differential operator \eqref{eq:intro_12}, in one spatial dimension $(d=1)$,  admits a family of self-adjoint realizations parametrized by the angle $\theta\in\n{S}^1$ (see Theorem \ref{theo:self_ext1D}). As a consequence,  the domains $\s{C}^\infty_{\rm c}(\R)$ or  $\s{S}(\R)$ \emph{can not} be cores for $\Pi$ (in contrast to \cite[Proposition  4.2.3]{amrein-boutet-georgescu-96}).
However, it turns out that all these self-adjoint realizations $\Pi_\theta$ are equivalent in the sense that there are unitary operators $L_\theta$ such that $\Pi_\theta=L_\theta\Pi_0L_\theta^*$. This fact immediately implies the independence of the spectrum by the   particular self-adjoint realization. In particular, it results that
the spectrum of every extension $\Pi_\theta$  is purely absolutely continuous and coincides with the real axis, \ie
\begin{equation}\label{eq:intro_14}
\sigma\big(\Pi_\theta\big)\;=\;\sigma_{\rm a.c.}\big(\Pi_\theta\big)\;=\;\R\;,\qquad\quad \forall \theta\in\n{S}^1.
\end{equation}

\medskip

We are now in position to state our first main result. Let us just recall that in dimension $d=1$ the only relevant parameter in the definition of the   thermal Hamiltonian is $\lambda>0$ since no  rotation $R_\gamma$ is required (\cf Note \ref{Note_01}). Then, according to Definition \ref{def:intro_01}, we can define the family of one-dimensional thermal Hamiltonians as
$$
H_{T,\theta}(\lambda)\;:=\;\lambda\;(\bb{F}S_\lambda) ^*\;\Pi_\theta\;(\bb{F}S_\lambda)\;,\qquad\quad \lambda>0,\;\;\theta\in\n{S}^{1}\;.
$$
In view of the unitary equivalence of the various realizations  $\Pi_\theta$ it follows that all the one-dimensional thermal Hamiltonians with a given coupling constant $\lambda>0$ are
 unitarily equivalent. For this reason we can focus on the special realization with $\theta=0$.

\begin{theorem}[Spectral theory  in $d=1$]\label{theo:main1}
For every $\lambda>0$, and up to a unitary equivalence, there is a unique one-dimensional thermal Hamiltonian on $L^2(\R)$ defined by
\begin{equation}\label{eq:intro_140}
H_T\;\equiv\;H_{T}(\lambda)\;:=\;\lambda\;(\bb{F}S_\lambda) ^*\;\Pi_0\;(\bb{F}S_\lambda)\;.
\end{equation}
The operator $H_{T}$ is self-adjoint on its domain 
$\s{D}(H_{T}):=(\bb{F}S_\lambda)^*[\s{D}(\Pi_0)]$
and has purely absolutely continuous spectrum given by 
$$
\sigma\big(H_{T}\big)\;=\;\sigma_{\rm a.c.}\big(H_{T}\big)\;=\;\R\;
$$
independently of $\lambda>0$.
\end{theorem}

\medskip

The proof of  Theorem \eqref{theo:main1} is a corollary Theorem \ref{theo:self_ext1D} and of Definition \ref{def:intro_01}. For the  determionation of the spectrum  one uses the invariance of the spectrum under unitary equivalences and the  \emph{spectral mapping theorem}.

\medskip

The operator $H_T$, defined by \eqref{eq:intro_140}, will be called  the \emph{standard realization} of the one-dimensional thermal Hamiltonian (with coupling constant $\lambda>0$).
Theorem \ref{theo:main1} expresses the fact that in dimension $d=1$ 
there is a  \virg{unique}  thermal Hamiltonian, 
at least in the sense that all relevant \emph{physical quantities}, which by definition must be invariant under unitary equivalences, can be calculated from $H_T$.

\medskip

Theorem \ref{theo:main1} can be complemented with some more precise information. First of all, it is possible to have a precise description of the domain $\s{D}(H_{T})$ (\cf Section \ref{sect:H_T1D-domain}). Let
\begin{equation}\label{eq:dom_pos}
\s{Q}(\R)\;:=\;\left\{\psi\in L^2(\R)\;\Big|\; \int_\R\dd x\; x^2|\psi(x)|^2\;<\;+\infty\right\}
\end{equation}
be the natural domain of the position operator. 
Let 
\begin{equation}\label{eq:int_ker_B_00}
(B_\lambda\psi)(x)\;:=\;\int_\R\dd y\; \bb{B}\left(x+\frac{1}{\lambda},y\right)\psi(y)
\end{equation}
be the unitary operator with integral kernel
\begin{equation}\label{eq:int_ker_B}
\bb{B}(x,y)\;:=\;\ii\frac{{\rm sgn}\left(x\right)-{\rm sgn}(y)}{2}\;J_0\left(2\sqrt{\left|xy\right|}\right)
\end{equation}
where
$$
{\rm sgn}(x)\;:=\;\left\{
\begin{aligned}
&\frac{x}{|x|}&&\text{if}\;\; x\neq0\\
&0&&\text{if}\;\; x=0
\end{aligned}
\right.
$$
is the  \emph{sign function} and $J_0$ is the 0-th Bessel function of the first kind\footnote{The kernel \eqref{eq:int_ker_B} is reminiscent of the Hankel transform of order $0$-th . This aspect is briefly discussed in Section \ref{sect:hank_tras}.} \cite{gradshteyn-ryzhik-07}. Then, it holds true that 
$$
\s{D}(H_{T})\;=\;B_\lambda[\s{Q}(\R)]\;.
$$
Moreover $\s{D}(H_{T})$ contains a dense core for $H_{T}$ given by
$$
\begin{aligned}
\s{D}_0(H_{T})\;:&=\;\s{S}(\R)\;+\;\C[\kappa_0]\\
&=\;\left\{\varphi\in L^2(\R)\;\Big|\varphi=\psi+ c \kappa_0 \;,\psi\in\s{S}(\R)\;, c\in\C\right\}
\end{aligned}
$$
and on this core $H_{T}$ acts according to (\cf Proposition \ref{prop:spec_fun_xi})
\begin{equation}\label{eq:act_HT_01}
\big(H_T\left(\psi+ c\kappa_0\right)\big)(x)\;:=\;-(1+\lambda x)\psi''(x)\;-\;\lambda\psi'(x)\;+\;c\kappa_1(x)\;
\end{equation}
where the (normalized) functions $\kappa_0$ and $\kappa_1$ are explicitly given by 
\begin{equation}\label{eq:act_HT_02}
\begin{aligned}
\kappa_0(x)\;:&=\;-\sqrt{\frac{8}{\pi}}\;{\rm sgn}\left(x+\frac{1}{\lambda}\right)\;{\rm kei}\left(2\sqrt{\left|x+\frac{1}{\lambda}\right|} \right)\\
\kappa_1(x)\;:&=\;\sqrt{\frac{8}{\pi}}\;{\rm ker}\left(2\sqrt{\left|x+\frac{1}{\lambda}\right|} \right)
\end{aligned}
\end{equation}
and ${\rm kei}(x)$ and ${\rm ker}(x)$ are the \emph{irregular Kelvin functions} of $0$-th order (see Section \ref{sec:irr_kelvin} and references therein). It is worth noting that the function  $\kappa_0$ introduces a jump discontinuity  around the \emph{critical point} $x_{\rm c}=-\lambda^{-1}$. The Hamiltonian $H_T$, acting on
  $\kappa_0$,  produces the wavefunction
   $\kappa_1$ which shows a logarithmic divergence around  $x_{\rm c}$. A similar singular behavior around the {critical point} $x_{\rm c}$ is detectable also in the
classical dynamics (\cf Section \ref{sect:App_classic}).

\medskip

The unitary propagator $U_T(t):=\expo{-\ii t H_T}$ acts as an integral operator
\begin{equation}\label{eq:int_ker_intro_UT}
(U_T(t)\psi)(x)\;:=\;\int_\R\dd y\; \bb{U}_{\lambda t}\left(x+\frac{1}{\lambda},y+\frac{1}{\lambda}\right)\psi(y)\;,\quad t\in\R\setminus\{0\}
\end{equation}
with kernel given by (\cf Proposition \ref{lemma:U_propag})
\begin{equation}\label{eq:int_ker_intro_UT2}
\bb{U}_\tau(x,y)\;:=\;\frac{{\rm sgn}(x)+{\rm sgn}(y)}{\ii 2\tau}\;\expo{\ii\frac{(x+y)}{\tau}}\;J_0\left(\frac{2}{\tau}\sqrt{\left|xy\right|}\right)
\end{equation}
for all $\tau\in\R\setminus\{0\}$.
Finally, the knowledge of the unitary propagator 
allows to compute the resolvent
$$
R_\zeta(H_T)\;:=\;(H_T-\zeta{\bf 1})^{-1}\;,\qquad \zeta\in\C\setminus\R
$$
by means of the Laplace transformation (see Section \ref{sect:TH_op_1D-case_res}).
It turns out that also $R_\zeta(H_T)$ is an integral operator
\begin{equation}\label{eq:int_ker_intro_RT}
(R_\zeta(H_T)\psi)(x)\;:=\;\frac{1}{\lambda}\int_\R\dd y\; \bb{Z}_{\frac{\zeta}{\lambda}}\left(x+\frac{1}{\lambda},y+\frac{1}{\lambda}\right)\psi(y)\;,
\end{equation}
with kernel $\bb{Z}_\alpha(x,y)$ given explicitly 
 by the (long) formulas \eqref{eq:impl_int_ker_res} and \eqref{eq:GOX_03}.

\medskip

Theorem \eqref{theo:main1} provides also the first step for the one-dimensional scattering theory of the thermal Hamiltonian. Indeed, one infers from  Theorem \eqref{theo:main1}  that $H_T$ does not admit bounded states and so generate a \virg{free-like} dynamics. In this work only the scattering theory  for a  special type of \emph{convolution} perturbations is discussed. The scattering theory for (physical) perturbations given by 
gravitational-matter potentials of type \eqref{eq:intro_07} presents several technical complications and will be treated in a separated work. By convolution perturbation we mean an integral operator $W_g$ acting on $\psi\in L^2(\R)$ as 
\begin{equation}\label{eq:scat_int_01}
(W_g\psi)(x)\;:=\;\int_\R\dd y\; g(x-y)\psi(y)
\end{equation}
where the kernel is chosen as $g\in L^1(\R)$.
Let us denote by
\begin{equation}\label{eq:scat_int_02}
H_{T,g}\;:=\;H_T\;+\;W_g
\end{equation}
the perturbed operator. As usual the wave operators for the pair $(H_T,H_{T,g})$ are defined by
\begin{equation}\label{eq:scat_int_03}
\Omega_g^{\pm}\;:=\;{\rm s}-\lim_{t\to\pm}\expo{\ii t H_{T,g}}\; \expo{-\ii t H_{T} }
\end{equation}
where the limit is meant in the strong sense.
In Section \ref{sect:scat_conv_pot} the following result will be proven.

\begin{theorem}[Scattering theory for convolution perturbations in $d=1$]\label{theo:main2}
Let $g\in L^1(\R)$ and $W_g$ the associated convolution perturbation defined by \eqref{eq:scat_int_01}. Then:
\begin{itemize}
\item[(i)] The perturbed operator $H_{T,g}$ defined by \eqref{eq:scat_int_02} is self-adjoint with domain $\s{D}(H_T)$ and
$$
\sigma\big(H_{T,g}\big)\;=\;\sigma_{\rm a.c.}\big(H_{T,g}\big)\;=\;\R\;.
$$
\end{itemize}
Let $\hat{g}$ be the Fourier transform of $g$ and assume that there are constants $\varepsilon>0$ and $C>0$ such that $|\hat{g}(x)|\leqslant Cx^{\frac{3}{2}}$ for all $|x|<\varepsilon$. Then: 
\begin{itemize}
\item[(ii)] The wave operators $\Omega_g^{\pm}
$ defined by \eqref{eq:scat_int_03} exist and are complete;
\vspace{1mm}
\item[(iii)] The S-matrix $S_g:=(\Omega_g^{+})^*\Omega_g^{-}$ is a constant phase given by
$$
S_g\;=\;\expo{-\ii\frac{\sqrt{2\pi}}{\lambda}\int_\R\dd s\; \frac{\hat{g}\left(s\right)}{s^2}}\;.
$$
\end{itemize}
\end{theorem}

%-----------------%
\medskip
\noindent
{\bf Structure of the paper.}
Section \ref{sect:aux_op_1D} is devoted to the study of the spectral theory of the auxiliary operator
$\Pi$ in the one-dimensional case. The spectral theory of the one-dimensional thermal Hamiltonian  $H_T$ is discussed in Section \ref{sect:TH_op_1D_spec} along with a subsection on the scattering theory by a convolution-type potentials. The classical dynamics of the thermal Hamiltonian (in any dimension) is studied in Section \ref{sect:App_classic}. Finally Appendix \ref{sect:app_spec_p}
 and Appendix \ref{app:tech_tool} 
contain some review material and some technical computations needed to make the present  work  self contained.

%-----------------%
\medskip
\noindent
{\bf Acknowledgements.} 
GD's research is supported
 by
the  grant \emph{Fondecyt Regular} -  1190204.
GD and VL are indebted to Claudio Fern\'andez, Marius M\u{a}ntoiu and Serge Richard for many stimulating discussions.

%--------------------%
%--------------------%
\section{The spectral theory of the operator $\Pi$}
\label{sect:aux_op_1D}
We already know from the general discussion in Section \ref{sec:intro_gen} that the operator 
 $\Pi$ defined by
\eqref{eq:intro_12}  (or formally by \eqref{eq:intro_13}) is {symmetric} and in turn {closable}. Moreover, Proposition \ref{prop:int_01} ensures that $\Pi$ admits self-adjoint extensions. While, on the one hand, these results are valid in every dimension, in
this section we will  classify all the 
self-adjoint extensions of $\Pi$ in dimension $d=1$ and we will
describe the
the spectral theory for this family of operators.

%---%
\subsection{Equivalence with the momentum operator}\label{sect:equiv_mpm_op}
In dimensional $d=1$ the operator $\Pi$ is initially defined by
\begin{equation}\label{eq:1D_01}
\begin{aligned}
\big(\Pi\psi\big)(x)\;&=\;\ii\left[x\psi(x)\;+\;x^2\frac{\dd\psi}{\dd x}(x)\right]\\
&=\;\ii x\frac{\dd}{\dd x}\left[x\psi(x)\right]
\end{aligned}
\qquad\quad \psi\in \s{S}(\R)\;.
\end{equation}
The last equality allows us to  identify
$$
\Pi\;\equiv\;-xpx
$$ 
on sufficiently regular functions.
 
\medskip

 The operator \eqref{eq:1D_01}
  is symmetric, hence closable, and its closure (still denoted with $\Pi$) has domain
$\s{D}_{0}$ given by
\eqref{eq:intro_15}.
In order to give a more precise characterization of  $\s{D}_{0}$ we will benefit from the transformation
$$
(I\psi)(x)\;:=\;\frac{1}{x}\psi\left(\frac{1}{x}\right)\;,\qquad\quad\psi\in L^2(\R)
\;.
$$
\begin{lemma}
$I$ is a unitary involution.
\end{lemma}
\proof
A direct computation shows that
$$
\begin{aligned}
||I\psi||^2\;&=\;\int_{\R}\frac{\dd x}{x^2}\left|\psi\left(\frac{1}{x}\right)\right|^2\;\;=\;-\int_{-\infty}^{+\infty}\dd \left(\frac{1}{x}\right)\left|\psi\left(\frac{1}{x}\right)\right|^2\\
&=\;-\int_{+\infty}^{-\infty}\dd s\left|\psi\left(s\right)\right|^2\;=\;||\psi||^2\;.
\end{aligned}
$$
Then $I$, initially defined on every \virg{good enough} dense domain, extends to an isometry on the whole $L^2(\R)$. From its very definition, it follows that $I^2\psi=\psi$. This shows that $I$ is an involution, and  in particular it is  invertible. As a consequence $I$ is also unitary.
\qed

\medskip

Instead of  $\Pi$ let us consider the transformed operator 
\begin{equation}\label{eq:1trans_mpm}
\wp\;:=\;I\;\Pi\;I
\end{equation} 
defined on the domain $\s{D}(\wp):=I[\s{D}_{0}]$. We use the standard notation $H^k(\Omega):=W^{k,2}(\Omega)\subset L^2(\Omega)$ for the $k$-th Sobolev space\footnote{For  the theory of Sobolev spaces we refer the reader to \cite[Chapter 8 \& Chapter 9]{brezis-87}.} with respect to the open set $\Omega\subseteq\R$.
Let 
$$
H^1_0(\R)\;:=\;\left\{\phi\in H^1(\R)\; \big|\; \phi(0)=0 \right\}
$$
be the space of the Sobolev functions on $\R$ vanishing in $x=0$. Let us point out that the latter requirement  makes sense since
Sobolev functions on $\R$ are uniquely identifiable with continuous functions \cite[Theorem 8.2]{brezis-87}. In view of this remark we will tacitly identify Sobolev functions with their continuous representative so that the following  inclusions $H^1_0(\R)\subset H^1(\R)\subset \s{C}(\R)$ hold.
\begin{proposition}\label{prop:pseudo_p}
The closed symmetric operator $\wp$ defined by \eqref{eq:1trans_mpm} coincides with the momentum operator on $H^1_0(\R)$, namely
$$
(\wp\phi)(x)\;=\;-\ii \phi'(x)\;,\qquad\quad \phi\in\s{D}(\wp)\;=\;H^1_0(\R)
$$
where $\phi'$ is the weak derivative of $\phi$.
\end{proposition}
\proof
The unitarity of $I$ implies that the graph norms of $\wp$ and $\Pi$ are related by
$||\phi||_{\wp}=||I\phi||_{\Pi}$ for all $\phi\in \s{D}(\wp)$. This gives
$$
\s{D}(\wp)\;=\;I\left[\overline{\s{S}(\R)}^{\;||\; ||_{\Pi}}\right]\;=\;
\overline{I[\s{S}(\R)]}^{\;||\; ||_{\wp}}\;.
$$
Let  $\phi\in I[\s{S}(\R)]$. Since $I\phi\in \s{S}(\R)$, one infers from \eqref{eq:1D_01} that
$$
(\Pi I\phi)(x)\;=\; \ii x\frac{\dd}{\dd x}\left[x(I\phi)(x)\right]
\;=\; \ii x\frac{\dd}{\dd x}\left[\phi\left(\frac{1}{x}\right)\right]\;=\;-\frac{\ii}{x}\frac{\dd\phi}{\dd x}\left(\frac{1}{x}\right)\;.
$$
Therefore
$$
(\wp\phi)(x)\;=\;(I(\Pi I\phi))(x)\;=\;-\ii\frac{\dd\phi}{\dd x}(x)
$$
acts as the momentum operator on $I[\s{S}(\R)]$. This implies that the domain of the closed operator $\wp$ is given by the closure of $I[\s{S}(\R)]$ with respect the Sobolev norm $
||\phi||^2_{H^1}:=||\phi||^2+||\phi'||^2$.
Let $\s{C}^\infty_{\rm c}(\R\setminus \{0\})$
be the set of smooth functions
having compact support separated from the origin. It holds true  that
\begin{equation}\label{eq:sob_inclus}
\s{C}^\infty_{\rm c}(\R\setminus \{0\})\;\subset\;I[\s{S}(\R)]\;\subset\;H^1_0(\R)\;.
\end{equation}
Indeed, let $\psi \in C_c^\infty(\R\setminus\{0\})$ supported in $[-b,-a]\cup[a,b]$
and $\phi:=I\psi$.  A direct inspection
shows that $\phi$ is a smooth function supported in $[-a^{-1},-b^{-1}]\cup[b^{-1},a^{-1}]$. This allows to conclude that $I[\s{C}^\infty_{\rm c}(\R\setminus \{0\})]\subseteq \s{C}^\infty_{\rm c}(\R\setminus \{0\})$. By exploiting the involutive character of $I$ one gets
$I[\s{C}^\infty_{\rm c}(\R\setminus \{0\})]= \s{C}^\infty_{\rm c}(\R\setminus \{0\})\subset \s{S}(\R)$ and in turn $\s{C}^\infty_{\rm c}(\R\setminus \{0\})\subset I[\s{S}(\R)]$.
For the second inclusion let us take $\phi \in I(\s{S}(\R))$ so that $\phi(x)=x^{-1}\psi( x^{-1})$ for some $\psi \in \s{S}(\R)$. Clearly, $\phi$ is smooth in $\R\setminus\left\{0\right\}$ and extends to a smooth function on $\R$ such that $\phi^{(n)}(0)=0$ for all $n \in \N$.
In particular  $\phi\in H^1_0(\R)$,
implying the second inclusion $I[\s{S}(\R)]\subset H^1_0(\R)$.
To conclude the proof it is enough to show that
the  closure of the space $\s{C}^\infty_{\rm c}(\R\setminus \{0\})$ with respect to the Sboolev norm $||\;||_{H^1}$ is (identifiable with) $H^1_0(\R)$. Let $\R_+:=(0,+\infty)$ and
$\R_-:=(-\infty,0)$ and  observe that
\begin{equation}\label{eq:sobo_ident}
\begin{aligned}
\overline{\s{C}^\infty_{\rm c}(\R\setminus \{0\})}^{\;||\;||_{H^1}}\;&=\;\overline{\s{C}^\infty_{\rm c}(\R_-)}^{\;||\;||_{H^1}}\;\oplus\;\overline{\s{C}^\infty_{\rm c}(\R_+)}^{\;||\;||_{H^1}}\\
&=\;W^{1,2}_0(\R_-)\;\oplus\;W^{1,2}_0(\R_+)\;=\;H^1_0(\R)
\end{aligned}
\end{equation}
where the notation for $W^{1,2}_0(\Omega)$ was borrowed from \cite[Section 8.3]{brezis-87}.
The last equality in \eqref{eq:sobo_ident} is a consequence of the fact that every element of $W^{1,2}_0(\R_\pm)$ can be uniquely identified with a continuous function that vanishes on the boundary $x=0$
\cite[Theorem 8.12]{brezis-87}. The identification \eqref{eq:sobo_ident}, along with the double inclusion \eqref{eq:sob_inclus},  implies the desired result $\s{D}(\wp)=\overline{I[\s{S}(\R)]}^{\;||\; ||_{H^1}}=H^1_0(\R)$.
\qed

\medskip

The first consequence of Proposition \ref{prop:pseudo_p} is  a precise description of the domain of the closed operator $\Pi$, \ie
\begin{equation}\label{eq:corol:dom_T}
\s{D}_{0}\;=\;I[\s{D}(\wp)]\;=\;\left\{\psi\in L^2(\R)\ \Big|\ \psi(x)=\frac{1}{x}\phi\left(\frac{1}{x}\right),\;\phi\in H^1_0(\R)\right\}\;.
\end{equation}
Unlike the functions in $H^1_0(\R)$, the elements of the domain $\s{D}_{0}$ are generally not continuous and can show  singularities in $x=0$. An example is the function $
\phi(x):=(1+x^2)^{-\frac{1}{3}}\expo{-\frac{1}{x^2}}$ which is evidently an element of $H_0^1(\R)$. Its image $\psi(x):=(I\phi)(x)=(x^3+x)^{-\frac{1}{3}}\expo{-x^2}
 $
 is divergent in $x=0$. 
 On the other hand 
  elements of $\s{D}_{0}$  have a decay at infinity which is at least of order 1.
\begin{proposition}\label{corol:dom_T}
Let $\psi\in\s{D}_{\;0}$. Then it holds true that
$$
\lim_{|x|\to\infty}(x\psi(x))\;=\;0\;.
$$ \end{proposition}
\proof
The claim  follows from the characterization \eqref{eq:corol:dom_T} which provides
$$
\lim_{x\to\pm\infty}( x\psi(x))\;=\lim_{t\to 0^\pm}\phi(t)\;=\;\phi(0)\;=\;0\;.
$$
In the last equality, the continuity of  $\phi\in H^1_0(\R)$ is used.
\qed

%---%
\subsection{Classification of self-adjoint extensions}\label{sect:self_ext1D-case}
We are now in position to  study the self-adjoint realizations of $\Pi$. In view of the unitary transform $I$  this is the same of studing the self-adjoint realization of the \emph{singular} momentum operator $\wp$. The latter is a classical problem strongly related with the  study of singular delta interactions for one-dimensional Dirac operators  \cite{gesztesy-seba-87,benvegnu-dabrowski-94,carlone-malamud-posilicano-13} (see also \cite[Appendix J]{albeverio-gesztesy-hoegh-gesztesy-88}).

\begin{proposition}\label{prob:first_step_wp}
The closed symmetric operator $\wp$ has deficiency indices equal to 1. Therefore,  the self-adjoint extensions of $\wp$ are in one-to-one correspondence with the angles  $\theta\in\n{S}^1\simeq[0,2\pi)$. The self-adjoint extension $\wp_\theta$ has domain
$$
\s{D}(\wp_\theta)\;:=\;\left\{\varphi\in L^2(\R)\;\Big|\varphi=\phi+ c \eta_\theta,\;\;\phi\in H^1_0(\R),\;\; c\in\C\right\}
$$
where
$$
\eta_\theta(x)\;:=\;\expo{-|x|}\expo{\ii{\rm sgn}(x)\frac{\theta}{2}}
$$
and acts has
\begin{equation}\label{eq:self_PP_01}
\wp_\theta\left(\phi+ c \eta_\theta \right)\;:=\;-\ii\phi'\;+\;c\eta_{\theta+\pi}\;.
\end{equation}
Finally, $\wp_0$ agrees with the standard momentum operator $p$ with domain $H^1(\R)$.
\end{proposition}

\proof
Since $\s{C}^\infty_{\rm c}(\R\setminus \{0\})=\s{C}^\infty_{\rm c}(\R_-)\oplus\s{C}^\infty_{\rm c}(\R_+)$ is dense (with respect to the graph norm) in the domain of $\wp$,
a standard argument shows that the
  adjoint operator $\wp^*$ acts as the weak derivative on its domain $\s{D}(\wp^*):=H^1(\R_-)\oplus H^1(\R_+)$ (see \eg \cite[Section VII.2]{reed-simon-I}). The eigenvalue equations $\wp^*\phi_\pm=\pm\ii\phi_\pm$ for the deficiency subspaces correspond to the
differential equations $\phi'_\pm=\mp\phi_\pm$
which admit in $\s{D}(\wp^*)$ the unique (normalized) weak solutions
$$
\phi_+(x)\;:=\;
\left\{
\begin{aligned}
&\sqrt{2}\expo{-x}&&\text{if}\;\; x>0\\
&0&&\text{if}\;\; x<0
\end{aligned}
\right.\;,\quad
\phi_-(x)\;:=\;
\left\{
\begin{aligned}
&0&&\text{if}\;\; x>0\\
&\sqrt{2}\expo{+x}&&\text{if}\;\; x<0
\end{aligned}
\right.\;.
$$
According to the von Neumann's theory for  self-adjoint extensions (\cf \cite[Section X.1]{reed-simon-II}) one has that the self-adjoint extensions of $\wp$ are parametrized by the unitary maps from $\s{K}_+=\C[\phi_+]\simeq\C$ to $\s{K}_-=\C[\phi_-]\simeq\C$. The later are identified by the angle $\theta\in\n{S}^1\simeq[0,2\pi)$ according to $U_\theta \phi_+:=\expo{-\ii\theta} \psi_-$. From the general theory \cite[Theorem X.2]{reed-simon-II} one has that the domain
of the self-adjoint extension $\wp_\theta$ is made by functions of the type
$
\phi+c'(\phi_++\expo{-\ii\theta} \phi_-)=\phi+c\eta_\theta
$
with $\phi\in H^1_0(\R)$ and $c,c'\in\C$ suitable complex coefficients. The action of
$\wp_\theta$ on the elements of its domain is given by
$$\wp_\theta\big(\phi\;+\;c'(\phi_+\;+\;\expo{-\ii\theta} \phi_-)\big)\;=\;-\ii\phi'\;+\;\ii c'(\phi_+\;-\;\expo{-\ii\theta} \phi_-)
$$
which translates into equation \eqref{eq:self_PP_01} in terms of the function $\eta_\theta$. Evidently, the standard momentum operator $p$ is a self-adjoint extension of $\wp$ since $H^1_0(\R)\subset H^1(\R)$. This extension corresponds to $\wp_0$ in view of the fact that $\eta_0\in H^1(\R)$.
\qed

\medskip

Although the symmetric operator $\wp$ admits several self-adjoint realizations, all these realizations are  in a sense equivalent. To express this fact in a precise way
we need to  introduce the family of unitary operators $L_\theta$ defined by
$$
(L_\theta\psi)(x)\;:=\;\expo{\ii{\rm sgn}(x)\frac{\theta}{2}}\psi(x)\;,\qquad\quad\psi\in L^2(\R)\;.
$$
\begin{proposition}\label{prob:second_step_wp}
The unitary operators $L_\theta$ intertwine all the self-adjoint realizations of the operator $\wp$. More precisely one has that
$$
\wp_\theta\;=\;L_\theta p L_\theta^*\;,\qquad\quad \theta\in\n{S}^1
$$
where $p=\wp_0$ is the standard momentum operator. As a consequence one has that
$$
\sigma(\wp_\theta)\;=\;\sigma_{\rm a.c.}(\wp_\theta)\;=\;\R\;,\qquad\quad \forall\;\theta\in\n{S}^1\;.
$$
\end{proposition}

\proof
Let us shows that $\s{D}(\wp_\theta)=L_\theta[H^1(\R)]$.
Every $\psi\in H^1(\R)$ can be decomposed as 
$(\psi-\psi(0)\eta_0)+\psi(0)\eta_0$. Evidently
$\phi:=\psi-\psi(0)\eta_0\in H^1_0(\R)$, $L_\theta\phi\in H^1_0(\R)$  and $L_\theta \eta_0=\eta_\theta$. Therefore, $L_\theta\psi\in \s{D}(\wp_\theta)$ which implies $L_\theta[H^1(\R)]\subseteq \s{D}(\wp_\theta)$. On the other hand every $\varphi\in  \s{D}(\wp_\theta)$ can be decomposed as $\varphi=L_\theta(L_{-\theta}\phi+c\eta_0)$ whit 
$L_{-\theta}\phi\in H^1_0(\R)$, and in turn $(L_{-\theta}\phi+c\eta_0)\in H^1(\R)$ proving the inverse inclusion $\s{D}(\wp_\theta)\subseteq L_\theta[H^1(\R)]$. Now, let  $\varphi\in  \s{D}(\wp_\theta)$. By exploiting the decomposition used above one has
$$
(L_\theta p L_\theta^*)\varphi\;=\;L_\theta(pL_{-\theta}\phi+cp\eta_0)\;=\;-\ii\phi'+c\eta_{\theta+\pi}
$$
where we used $(p\eta_0)(x)=\eta_0(x)\expo{\ii{\rm sgn}(x)\frac{\pi}{2}}$ and $pL_{-\theta}\phi=L_{-\theta}p\phi$ in view of $\phi\in H^1_0(\R)$. Hence, a comparison with \eqref{eq:self_PP_01} shows that $L_\theta p L_\theta^*=\wp_\theta$ on the domain $ \s{D}(\wp_\theta)$.
\qed

\begin{remark}
{\upshape
The unitary equivalence of the different realizations  $\wp_\theta$ can be understood in terms of the celebrated \emph{Stone-von Neumann theorem} (see \eg \cite{rosenberg-04}). Indeed, a direct computation shows that
$$
(x\;\wp_\theta\;-\; \wp_\theta\;x)\varphi\;=\;\ii\varphi\;,\qquad\quad\varphi\in \s{C}^\infty_{\rm c}(\R\setminus \{0\})
$$
and $\s{C}^\infty_{\rm c}(\R\setminus \{0\})=\s{C}^\infty_{\rm c}(\R_-)\oplus\s{C}^\infty_{\rm c}(\R_+)$ is dense in $L^2(\R_-)\oplus L^2(\R_+)=L^2(\R)$. Therefore, by continuous extension, one can unambiguously define the commutation relation $[x,\wp_\theta]=\ii{\bf 1}$ which means that the pair $(x,\wp_\theta)$ satisfies  the \emph{canonical commutation relation}. As a result, the Stone-von Neumann theorem ensures that $\wp_\theta$ is unitarily equivalent to the standard momentum $p$.}
   \hfill $\blacktriangleleft$
\end{remark}

\medskip

Proposition \ref{prob:first_step_wp} 
 provides the key result for  the complete description of the self-adjoint extensions of $\Pi$.
\begin{theorem}[Self-adjoint extensions: one-dimensional case]
\label{theo:self_ext1D}
The self-adjoint extensions of the closed symmetric operator $\Pi$ initially defined by \eqref{eq:1D_01} are in one-to-one correspondence with the angles  $\theta\in\n{S}^1$. The self-adjoint extension $\Pi_\theta$ has domain
$$
\begin{aligned}
\s{D}(\Pi_\theta)\;:&=\;\left\{\varphi\in L^2(\R)\;\Big|\varphi=\psi+ c \zeta_\theta \;,\psi\in\s{D}_{\;0}\;, c\in\C\right\}
\end{aligned}
$$
where
$$
\zeta_\theta(x)\;:=\;
\frac{1}{x}\expo{-\frac{1}{|x|}}\expo{\ii{\rm sgn}(x)\frac{\theta}{2}}
$$
and acts has
$$
\Pi_\theta\left(\psi+ c\zeta_\theta \right)\;:=\;\Pi\psi\;+\;c\zeta_{\theta+\pi}\;.
$$
All the self-adjoint realizations are unitarily equivalent, \ie $\Pi_\theta\;=\;L_\theta \Pi_0 L_\theta^*$ for all $\theta\in\n{S}^1$.
Finally one has that
$$
\sigma(\Pi_\theta)\;=\;\sigma_{\rm a.c.}(\Pi_\theta)\;=\;\R\;,\qquad\quad \forall\;\theta\in\n{S}^1\;.
$$
\end{theorem}
\proof
This  is a direct consequence of the unitary equivalence established in Proposition \ref{prop:pseudo_p} which allows to define the  self-adjoint realizations of $\Pi$ by
$\Pi_\theta:=I\wp_\theta I$. 
Therefore,
 the statement is nothing more than
a rephrasing of Proposition \ref{prob:first_step_wp} 
 and Proposition \ref{prob:second_step_wp}. The formula $\Pi_\theta\;=\;L_\theta \Pi_0 L_\theta^*$ is justified by the commutation relation $L_\theta I=I L_\theta$.
\qed

\medskip

In view of the unitary equivalence among all the self-adjoint realizations of $\Pi$  we can focus the attention only in  a \virg{preferred} realization.

\begin{definition}[Standard realization]
We will call $\Pi_0=\Pi_{\theta=0}$  the \emph{standard
} self-adjoint realization of the operator initially defined by \eqref{eq:1D_01}. 
\end{definition}

%---%
\subsection{Boundary triplets}\label{sect:bound_trip}
The problem of the determination of the self-adjoint realizations of $\wp$ or $\Pi$
can be  investigated also with the theory of the \emph{boundary triplets} \cite[Chapter 14]{schmudgen-12}.
Let us start with the operator $\wp$ and its adjoint $\wp^*$. According to \cite[Definition 14.2]{schmudgen-12}, a boundary triplet for $\wp^*$ is a triplet $(\n{H},\Gamma_0,\Gamma_1)$
made by an Hilbert space $\n{H}$ and linear maps $\Gamma_0,\Gamma_1$ from $\s{D}(\wp^*)$ to $\n{H}$ that satisfy the \emph{abstract Green's identity}
$$
\langle\wp^*\varphi,\psi\rangle\;-\;\langle\varphi,\wp^*\psi\rangle\;=\;\langle\Gamma_0\varphi,\Gamma_1\psi\rangle_{\n{H}}\;-\;\langle\Gamma_1\varphi,\Gamma_0\psi\rangle_{\n{H}}\;,\quad\forall\; \varphi,\psi\in\s{D}(\wp^*)
$$  
and the mapping $\s{D}(\wp^*)\ni\varphi\mapsto(\Gamma_0\varphi,\Gamma_1\varphi)\in \n{H}\times\n{H}$ is surjective. Since  the operator $\wp^*$ acts as the weak derivative on its domain 
$$
\s{D}(\wp^*)\;:=\;H^1(\R_-)\;\oplus\; H^1(\R_+)\;,
$$ 
an integration by parts provides 
$$
\langle\wp^*\varphi,\psi\rangle-\langle\varphi,\wp^*\psi\rangle\;=\;\ii\left(\overline{\varphi(0^-)}\psi(0^-)-\overline{\varphi(0^+)}\psi(0^+)\right)
$$
where $\varphi(0^\pm):=\lim_{x\to0^\pm}\varphi(x)$ and similarly for $\psi(0^\pm)$.
A comparison with the abstract Green's identity shows that the triplet $(\n{H},\Gamma_0,\Gamma_1)$ can be fixed in the following way: $\n{H}:=\C$;
$$
\Gamma_0\varphi\;:=\;\frac{\varphi(0^+)-\varphi(0^-)}{\ii\sqrt{2}}\;,\qquad \Gamma_1\varphi\;:=\;\frac{\varphi(0^+)+\varphi(0^-)}{\sqrt{2}}\;.
$$
The surjectivity condition is obviously satisfied. Observe that ${\rm Ker (\Gamma_0)}\cap{\rm Ker (\Gamma_1)}=H^1_0(\R)=\s{D}(\wp)$.
The self-adjoint extensions of $\wp$ are in one-to-one correspondence with the self-adjoint operators on $\n{H}=\C$ \cite[Theorem 14.10]{schmudgen-12}. More precisely, the self-adjoint extensions of $\wp$ can be parametrized 
by a real number $\gamma\in\R\cup\{\infty\}$ 
which defines a restriction $\wp_\gamma:=\wp^*|_{\s{D}_\gamma}$ where the domain $\s{D}_\gamma\subset \s{D}(\wp^*)$ is defined by
\begin{equation}\label{eq:bound_cond}
\begin{aligned}
{\s{D}_\gamma}\;:&=\;\left\{\varphi\in \s{D}(\wp^*)\;\Big|\gamma\Gamma_0\varphi=\Gamma_1\varphi\right\}\\
&=\;\left\{\varphi\in \s{D}(\wp^*)\;\Big|\expo{-\ii\arctan\left(\frac{1}{\gamma}\right)}\varphi(0^+)=\expo{\ii\arctan\left(\frac{1}{\gamma}\right)}\varphi(0^-)\right\}\;.
\end{aligned}
\end{equation}
A comparison with  Proposition \ref{prob:first_step_wp} shows that the self-adjoint extensions $\wp_\theta$ and $\wp_\gamma$ are related by the equation $\theta(\gamma)=\arctan\left(\frac{1}{\gamma}\right)$. In particular, the standard momentum is identified by $\gamma=\infty$ which corresponds to $\theta=0$. 
The definition \eqref{eq:bound_cond} provides the description of the domain of 
$\wp_{\theta}$ in therms of \emph{boundary conditions}. The same can be done for the the self-adjoint extensions $\Pi_\theta$ with the help of the unitary operator $I$.
 A direct computation shows that 
$$
\begin{aligned}
\s{D}(\Pi_\theta)\;:&=\;\left\{\varphi\in I[\s{D}(\wp^*)]\;\Big|\;\expo{-\ii\frac{\theta}{2}}(x\varphi)(+\infty)=\expo{+\ii\frac{\theta}{2}}(x\varphi)(-\infty)\right\}
\end{aligned}
$$
where $(x\varphi)(\pm\infty):=\lim_{x\to\pm\infty}x\varphi(x)$.

%%%%
\subsection{Unitary propagator}\label{sect:1D-case_up}
Let
\begin{equation}\label{eq:unit_prop}
V_\theta(t)\;:=\;\expo{-\ii t \Pi_\theta}\;,\qquad\quad t\in\R
\end{equation}
be the \emph{unitary propagator} defined by the self-adjoint operator $\Pi_\theta$ on $L^2(\R)$. The description of $V_\theta(t)$ is provided in the following theorem.
\begin{theorem}
Let $V_\theta(t)$ be the unitary group defined by \eqref{eq:unit_prop}. It holds true that
$$
\big(V_\theta(t)\psi\big)(t)\;=\;\frac{\expo{\ii\frac{\theta}{2} \big(1-{\rm sgn}(1-tx)\big){\rm sgn}(x)}}{1-t x}\;\psi\left(\frac{x}{1-t x}\right)\;,\qquad \psi\in L^2(\R)\;.
$$
\end{theorem}
\proof
We can use the unitary equivalence $\Pi_\theta=IL_\theta p L_\theta^*I$ proved in Section \ref{sect:self_ext1D-case}. This implies that $V_\theta(t)=IL_\theta \expo{-\ii t p} L_\theta^*I$ along with the well-known fact $(\expo{-\ii t p} \psi)(x)=\psi(x-t)$. The proof of the claim follows by  a direct computation.
\qed

\medskip

For each $t\in\R$ let us consider the map $f_t:\R \cup \{\infty\} \to \R \cup \{\infty\}$
defined by
\begin{equation}\label{eq:flow}
f_{t}(x)\;:=\;
\left\{
\begin{aligned}
&\frac{x}{1-t x}&\text{if}\;\;&x\in \R \setminus \{t^{-1}\}\\
&\infty&\text{if}\;\;&x=t^{-1}\\
&-t^{-1}&\text{if}\;\;&x=\infty\;.\\
\end{aligned}
\right.
\end{equation}
with the convention that $\pm 0^{-1}\equiv\infty$.
The family of these maps defines an \emph{$\R$-flow} in the sense that the following relations hold:
\begin{equation}
\left\{
\begin{aligned}
&f_0\;=\;{\rm Id}\\
&f_{t_1}\circ f_{t_2}\;=\; f_{t_1+t_2}\\
&f_t^{-1}\;=\;f_{-t}\;\\
\end{aligned}\right.\qquad\quad \forall\; t,t_1,t_2\in\ \R \;.
\end{equation}
The flow $f_t$ allows to rewrite  the action of $V_\theta(t)$ in the form 
\begin{equation}\label{eq:flow01}
\big(V_\theta(t)\psi\big)(t)\;=\;\expo{\frac{\ii}{2} \big(1-{\rm sgn}(1-tx)\big)\big({\rm sgn}(x)\theta+\pi\big)}\;\sqrt{(\partial_xf_t)(x)}\;\psi\left(f_t(x)\right)\;.
\end{equation}
When $\theta=\pi$ the exponential prefactor is 1 and equation \eqref{eq:flow01} agrees with the definition of the \emph{$C_0$-group} associated to the flow $f_t$ as defined in  \cite[Section 4.2]{amrein-boutet-georgescu-96}. It is interesting to notice that the flow $f_t$ is not of class $\s{C}^\infty$ and the generator of the flow
$$
F(x)\;:=\;\left.\frac{\dd f_t}{\dd t}\right|_{t=0}(x)\;=\;x^2
$$
has an unbounded first derivative. Therefore the flow $f_t$  doesn't meet the conditions of  \cite[Lemma 4.2.2 \& Proposition 4.2.3]{amrein-boutet-georgescu-96}.
The latter fact explains why \cite[Proposition 4.2.3]{amrein-boutet-georgescu-96} doesn't apply to 
 the operator $\Pi\equiv-\frac{1}{2}(p F(x)+F(x)p)$  which indeed is not essentially self-adjoint on $\s{C}^\infty_{\rm c}(\R)$.

%%%%
\subsection{Resolvent and Green function}\label{sect:1D-case_res}
The resolvent of the of the operator $\Pi_\theta$ can be derived from the resolvent of the standard momentum operator $p$ by exploiting the various unitary equivalences described in Section \ref{sect:self_ext1D-case}. For every $\zeta\in\C\setminus\R$ the 
resolvent of  $\Pi_\theta$ at $\zeta$ is 
defined as
\begin{equation}
R_\zeta(\Pi_\theta)\;:=\;\left(\Pi_\theta-\zeta{\bf 1}\right)^{-1}\;=\;
L_\theta I\left(p-\zeta{\bf 1}\right)^{-1}I L_\theta^*\;.
\end{equation}
The next results shows that $R_\zeta(\Pi_\theta)$ is an integral operator.
\begin{proposition}\label{prop:int_kernel}
Let $\zeta:=\epsilon\pm\ii\delta\in\C\setminus\R$ with $\delta>0$. The resolvent  $R_\zeta(\Pi_\theta)$ acts as
$$
\big(R_\zeta(\Pi_\theta)\psi\big)(x)\;=\;\int_{\R}\dd y\; \bb{R}^\theta_\zeta(x,y)\;\psi(y)\;,\qquad\quad \psi\in L^2(\R)
$$
with kernel given by
$$
\bb{R}^\theta_{\epsilon\pm\ii\delta}(x,y)\;:=\;\frac{\expo{\ii\big({\rm sgn}(x)-{\rm sgn}(y)\big)\frac{\theta}{2}}}{\mp\ii xy}\; \Theta\left(\pm\left(\frac{1}{x}-\frac{1}{y}\right)\right)\;\expo{\ii\epsilon\left(\frac{1}{x}-\frac{1}{y}\right)}\expo{-\delta\left|\frac{1}{x}-\frac{1}{y}\right|}
$$
where $\Theta$ is the \emph{Heaviside function}.
\footnote{The \emph{Heaviside function} is defined by
$
\Theta(x)\;:=\;\left\{
\begin{aligned}
&1&&\text{if}\;\; x>0\\
&\frac{1}{2}&&\text{if}\;\; x=0\\
&0&&\text{if}\;\; x<0\;.\\
\end{aligned}
\right.
$}
\end{proposition}
\proof
The integral kernel $\bb{R}^0_\zeta$ of the resolvent of $\Pi_0$ can be obtained
from the Green's function $\bb{G}^0_{\zeta}$ of the  standard momentum operator (see Appendix \ref{app:B1}). A direct computation provides
$$
\big(R_\zeta(\Pi_0)\psi\big)(x)\;=\;\big(I\left(p-\zeta{\bf 1}\right)^{-1}I\psi\big)(x)\;=\;\frac{1}{x}\int_{\R}\dd y\; \bb{G}^0_\zeta\left(\frac{1}{x},y\right)\;\frac{1}{y}\psi\left(\frac{1}{y}\right)\;.
$$
The explicit expression of $\bb{G}^0_{\zeta}$ given in \eqref{eq:green_p} and 
 a change of  variable in the integral 
provide
$$
\bb{R}^0_{\zeta}(x,y)\;:=\;\frac{1}{xy} \bb{G}^0_\zeta\left(\frac{1}{x},\frac{1}{y}\right)\;.
$$
Since $L_\theta$ is a multiplication operator, the relation between the kernels for $\theta=0$ and $\theta\neq 0$ is simply given by
$$
\bb{R}^\theta_{\zeta}(x,y)\;:=\;\expo{\ii\big({\rm sgn}(x)-{\rm sgn}(y)\big)\frac{\theta}{2}}\bb{R}^0_{\zeta}(x,y)\;=\;\frac{\expo{\ii\big({\rm sgn}(x)-{\rm sgn}(y)\big)\frac{\theta}{2}}}{xy} \bb{G}^0_\zeta\left(\frac{1}{x},\frac{1}{y}\right)\;.
$$
This concludes the proof.
\qed

\medskip

It is worth noting that that along the diagonal one has
$$
\bb{R}^\theta_{\zeta}(x,x)\;=\;{\rm sgn}({\rm Im}(\zeta))\;\frac{\ii}{2x^2}\;
$$
for all $\zeta\in\C\setminus\R$ and $\theta\in\n{S}^1$.

%--------------%
\subsection{Spectral
measure and density of states}\label{sect:1D-case_spec_meas}
Let $\mu_\psi^{\theta}$ be the spectral measure of the operator $\Pi_\theta$ associated with the normalized state   $\psi\in L^2(\R)$. We know from
Theorem \ref{theo:self_ext1D}  that $\Pi_\theta$ as a purely absolutely continuous spectrum
which coincides with $\R$.
This implies that the spectral  measure $\mu_\psi^{\theta}$
 is purely absolutely continuous. More precisely one has that
 $$
\mu_\psi^{\theta}(\dd\epsilon)\;:=\;f_\psi^\theta(\epsilon)\;\dd\epsilon 
$$
with $f_\psi^\theta\in L^1(\R)$ a non-negative function. The next result provides a description of $f_\psi^\theta$.
\begin{proposition}
Let $\mu_\psi^{\theta}$ be the spectral measure of the operator $\Pi_\theta$ associated with the (normalized) state   $\psi\in L^2(\R)$. Then $\mu_\psi^{\theta}$ is absolutely continuous with respect to the  
Lebesgue measure $\dd\epsilon$ in $\R$ and 
\begin{equation}\label{eq:spec_meas_01}
\mu_\psi^{\theta}(\dd\epsilon)\;:=\;|\widehat{\phi}_\theta(\epsilon)|^2\;\dd\epsilon
\end{equation}
where $\widehat{\phi}_\theta:=\bb{F}(\phi_\theta)$ is the Fourier transform of the function
$$
{\phi}_\theta(x)\;:=\;(L_\theta^*I\psi)(x)\;=\;\frac{\expo{-\ii{\rm sgn}(x)\frac{\theta}{2}}}{x}\psi\left(\frac{1}{x}\right)\;.
$$
\end{proposition}
\proof
From the unitary equivalence $\Pi_\theta=IL_\theta p L_\theta^*I$ one gets
$$
F^\theta_\psi(\zeta)\;:=\;\langle\psi,(\Pi_\theta-\zeta{\bf 1})^{-1} \psi\rangle\;=\;\langle\psi,IL_\theta(p-\zeta{\bf 1})^{-1} L_\theta^*I\psi\rangle\;=\;F^p_{\phi_\theta}(\zeta)\;.
$$
Following the arguments in Appendix \ref{app:B2}
on gets
$$
\lim_{\delta\to 0^+}\frac{1}{\pi}\;{\rm Im}\left(F_\psi^{\theta}(\epsilon+\ii\delta)\right)\;=\;f_{\phi_\theta}^p(\epsilon)\;=\;|\widehat{\phi}_\theta(\epsilon)|^2
$$ 
where the last equality is justified by \eqref{eq:spec_meas_p}.
This concludes the proof.
\qed

\medskip

In order to define the integrated density of states (IDOS) of $\Pi_\theta$ let us start  by introducing the spectral projections $P^\theta_\epsilon$ of $\Pi_\theta$  defined by
$$
P^\theta_\epsilon\;:=\;
\left\{
\begin{aligned}
&\chi_{[0,\epsilon]}(\Pi_\theta)&\text{if}\;\;& \epsilon>0\\
&\chi_{[\epsilon,0]}(\Pi_\theta)&\text{if}\;\;& \epsilon<0\;.\\
\end{aligned}
\right.
$$
Let $(Q_\Lambda\psi)(x)=\chi_{\Lambda}(x)\psi(x)$ be  the projection which restricts the functions $\psi\in L^2(\R)$  on the interval $\Lambda=[a,b]$. 
Let us introduce the function $\s{N}^\theta_\Lambda:\R\to\R$ defined by
\begin{equation}\label{eq:density_T_01}
\s{N}^\theta_\Lambda(\epsilon)\;:=\;\frac{{\rm sgn}(\epsilon)}{|\Lambda|}\;{\rm Tr}\left(P^\theta_\epsilon Q_\Lambda\right)\;.
\end{equation}
Definition \eqref{eq:density_T_01} is 
well posed in view of  the following result:
\begin{lemma}\label{lemma:dens_T}
Let $\Lambda:=[a,b]$ with $ab>0$.
The operator $P^\theta_\epsilon Q_\Lambda$ is trace class and 
$$
\s{N}^\theta_\Lambda(\epsilon)\;=\;\frac{1}{ab}\frac{\epsilon}{2\pi}\;
$$
independently of $\theta$.
\end{lemma}
\proof
By combining the spectral theorem with the unitary equivalence between $\Pi_\theta$ and $p$ one gets that $P^\theta_\epsilon=L_\theta IP_\epsilon I L_\theta^*$ where 
\begin{equation}\label{eq:density_T_02}
P_\epsilon\;:=\;
\left\{
\begin{aligned}
&\chi_{[0,\epsilon]}(p)&\text{if}\;\;& \epsilon>0\\
&\chi_{[\epsilon,0]}(p)&\text{if}\;\;& \epsilon<0\;.
\end{aligned}
\right.
\end{equation}
  This means that  $P^\theta_\epsilon Q_\Lambda=L_\theta (IP_\epsilon I Q_\Lambda)L_\theta^*=L_\theta I (P_\epsilon I Q_\Lambda I)IL_\theta$. Thus, to prove that  $P^\theta_\epsilon Q_\Lambda$ is trace-class 
it is sufficient to prove that $P_\epsilon (I Q_\Lambda I)$ is trace-class.  Let $b>a>0$ or $a<b<0$. A direct computation shows that
$$
(I Q_{[a,b]}I\psi)(x)\;=\;\chi_{[a,b]}\left(\frac{1}{x}\right)\psi(x)\;=\;\chi_{[b^{-1},a^{-1}]}\left(x\right)\psi(x)\;,\quad \psi\in L^2(\R)\;,
$$
namely $I Q_\Lambda I= Q_{\tilde{\Lambda}}$ with $\tilde{\Lambda}:=[b^{-1},a^{-1}]$. This implies that 
$P_\epsilon (I Q_\Lambda I)=P_\epsilon Q_{\tilde{\Lambda}}$  is trace-class in view of \cite[Theorem XI 20]{reed-simon-III}. Moreover, one has that
$$
\s{N}^\theta_\Lambda(\epsilon)\;=\;\frac{|\tilde{\Lambda}|}{|\Lambda|}\frac{{\rm sgn}(\epsilon)}{|\tilde{\Lambda}|}\;{\rm Tr}\left(P^\theta_\epsilon Q_\Lambda\right)\;=\;\frac{a^{-1}-b^{-1}}{b-a}\s{N}^p_{\tilde{\Lambda}}(\epsilon)
$$
where $\s{N}^p_{\tilde{\Lambda}}(\epsilon)$ is the local density of states for the operator $p$ in the region $\tilde{\Lambda}$. The proof follows by using Lemma \ref{lemma:IDOS_p_02}.
\qed

\medskip

The quantity $\s{N}^\theta_\Lambda(\epsilon)$ measures the volumetric density of states up to the energy $\epsilon$ localized in the region $\Lambda$. States with negative energy are counted as \virg{negative} states.  Lemma \ref{lemma:dens_T} shows that this number  is not homogeneous in space. One can ask  how this number changes for fixed volume in function of the spatial localization. Let $\ell>0$ and  set $\Lambda_{x,\ell}:=[x,x+\ell]$ when $x>0$ or $\Lambda_{x,\ell}:=[x-\ell,x]$ when $x<0$. Then
$$
\s{N}^\theta_{\Lambda_{x,\ell}}(\epsilon)\;:=\;\frac{1}{x^2+|x|\ell}\frac{\epsilon}{2\pi}\;.
$$
Since the density decreases as $x^{-2}$ in function of the spatial localization and as $\ell^{-1}$
in function of the volume one immediately concludes that the majority of states are concentrated around $x=0$ with a divergent density.

\medskip

Ultimately, the spatial 
inhomogeneity of $\s{N}^\theta_\Lambda$   is a consequence of the fact that $\Pi_\theta$
breaks the invariance under spatial translations. To define a density of states on the thermodynamic limit a precise prescription on how to carry out the spatial average is necessary. Let us define the \emph{principal value integral density of states} (pv-IDOS) as
$$
{\rm pv}-\s{N}^\theta(\epsilon)\;:=\;\lim_{L\to\infty}\frac{L}{2(L^2-1)}\;{\rm sgn}(\epsilon)\;{\rm Tr}\left(P_\epsilon Q'_L\right)
$$
where $Q'_L:= Q_{[-L,L]}-Q_{[-L^{-1},L^{-1}]}$. From Lemma \eqref{lemma:dens_T}
one immediately gets that
$$
{\rm pv}-\s{N}^\theta(\epsilon)\;:=\;\frac{\epsilon}{2\pi}\;.
$$

%--------------------%
%--------------------%
\section{The spectral theory of the thermal Hamiltonian}
\label{sect:TH_op_1D_spec}
The thermal Hamiltonian $H_T$ is defined by equation \eqref{eq:intro_140} as the conjugation of $\Pi_0$ through the unitary $\bb{F}S_\lambda$.
For this reason the spectral theory of $H_T$ (summarized by  Theorem \ref{theo:main1}) is equivalent to the spectral theory of $\Pi_0$ studied in Section \ref{sect:aux_op_1D}. The  next section is mainly devoted to the translation of the results obtained for $\Pi_0$ to $H_T$
by exploiting the precise form of the unitary
$\bb{F}S_\lambda$.
%%%%
\subsection{Description of the domain}\label{sect:H_T1D-domain}
By construction the domain of $H_T$ is given by
$$
\s{D}(H_T)\;:=\;(\bb{F}S_\lambda)^*[\s{D}(\Pi_0)]\;=\;({S}_\lambda^*{B} \bb{F}^*)[H^1(\R)]
$$
with ${B}:=\bb{F}^* I\bb{F}$. The last equality is justified by $\s{D}(\Pi_0)=I[\s{D}(\wp_0)]$ and $\s{D}(\wp_0)=H^1(\R)$. It is known that the Fourier transform of 
$H^1(\R)$ is the domain of the position operator \cite[Chapter IX]{reed-simon-II} defined by \eqref{eq:dom_pos}.
Therefore,  the domain $H_T$ is made by functions in $\s{Q}(\R)$ transformed by the operator ${S}_\lambda^* B$.
The operators ${B}$ and $B_\lambda:={S}_\lambda^*{B}$ have a description in terms of integral kernels.
\begin{lemma}\label{lemma:B_tild}
On the dense domain $L^2(\R)\cap L^1(\R)$
the operator ${B}=\bb{F}^*I\bb{F}$ acts as an integral operator with kernel given by \eqref{eq:int_ker_B}. As a consequence 
$B_\lambda:={S}_\lambda^*{B}$ acts according to 
\eqref{eq:int_ker_B_00}.
\end{lemma}
\proof
Let us start with the computation of the  kernel of $B$ acting on $\psi \in L^2(\R)\cap L^1(\R)$. 
Then $\bb{F}(\psi)\in L^2(\R)\cap\s{C}_0(\R)$, namely $\bb{F}(\psi)$ is a square-integrable continuous function that vanishes at infinity.  For every $n\in\N$, let $\chi_{I_{n}}$ be the characteristic function of the interval $I_{n}:=[-n,-n^{-1}]\cup[n^{-1},n]$
Since  $\bb{F}(\psi)-\bb{F}(\psi)\chi_{I_{n}}=\bb{F}(\psi)\chi_{I^c_{n}}$, where $I^c_{n}$ is the complement of $I_{n}$, one can prove that $\bb{F}(\psi)\chi_{I_{n}}\to \bb{F}(\psi)$ in the $L^2$-topology when $n\to+\infty$. Thus, the unitarity of the Fourier transform implies that 
$\psi_n\to \psi$ in the $L^2$-topology where $\psi_{n}:=\bb{F}^*(\bb{F}(\psi)\chi_{I_{n}})=\psi\ \ast \bb{F}^*(\chi_{I_{n}})$ and  $\ast$ denotes the convolution.
Since $B$ is a unitary operator one gets $B\psi_n\to B\psi$ with respect to the $L^2$-topology. 
An explicit computation provides
$$
\begin{aligned}
(B\psi_{n})(x)\;&=\;(\bb{F}^*I\bb{F}\psi_{n})(x)\\
 &=\;(\bb{F}^*I\bb{F}\left(\psi\ \ast \bb{F}^*\chi_{I_{n}})\right)(x)\\
  &=\;(\bb{F}^*I(\bb{F}\psi) \chi_{I_{n}})(x)\\
  & =\;\frac{1}{\sqrt{2\pi}}\int_{\R}\dd u\;\expo{\ii ux}\frac{1}{u}(\bb{F}\psi)\left(\frac{1}{u}\right) \chi_{I_{n}}\left(\frac{1}{u}\right)\\
& =\;\frac{1}{2\pi}\int_{I_{n}}\dd u\; \expo{\ii ux}\frac{1}{u}\left(\int_\R\dd y\; \expo{-\ii\frac{ y}{u}}\psi(y)\right)\;
\end{aligned}
$$
where in the last two equalities we used the fact that $I(\bb{F}\psi) \chi_{I_{n}}$ and $\psi$
are $L^1$-functions (this justifies the use of the integral representation of $\bb{F}$ and $\bb{F}^*$) and the equality $\chi_{I_{n}}(u^{-1})=\chi_{I_{n}}(u)$. 
Since the function
$g_x(y,u):=\frac{1}{u}\expo{\ii xu}\expo{\frac{-\ii y}{u}}\psi(y)$ is absolutely integrable in $\R \times I_{n}$ one can invoke the Fubini-Tonelli theorem to change the order of integration. This provides
\begin{equation}\label{eq:misck_01}
 (B\psi_{n})(x)\;=\;\frac{1}{2\pi}\int_\R\dd y\; \psi(y)\left(\int_{I_{n}}\dd u\; \frac{\expo{\ii xu}\expo{-\ii \frac{y}{u}}}{ u}\right)\;.
\end{equation}
Corollary \ref{corol:kern_main} says that 
$$
\lim_{n\to \infty}\int_{I_{n}}\dd u\; \frac{\expo{\ii xu}\expo{-\ii \frac{y}{u}}}{ u}\;=\;2\pi\;\bb{B}(x,y)\;.
$$
And
$$
\left|\int_{I_{n}}\dd u\; \frac{\expo{\ii xu}\expo{-\ii \frac{y}{u}}}{ u}\right|\;\leqslant\;4\pi
$$ 
for all $n>n_0$. In view of the bound above one can use the 
Lebesgue's dominated convergence theorem in \eqref{eq:misck_01} providing the formula
\begin{equation}\label{eq:misck_02}
\lim_{n\to+\infty} (B\psi_{n})(x)\;=\;\int_{-\infty}^{+\infty}\dd y\; \bb{B}(x,y)\;\psi(y)\;.
\end{equation}
Equation \eqref{eq:misck_02} says that $B\psi_{n}$ converges pointwise to the 
integral in the right-hand side. Since 
$B\psi_{n}$ converges to $B\psi$ in the $L^2$-topology it follows there exists a subsequence 
$B\psi_{n_k}$ which converges pointwise (almost everywhere) to $B\psi$ \cite[Theorem 4.9 (a)]{brezis-87}. Then the unicity of the limit assures that 
$B\psi$ coincides with the right-hand side of \eqref{eq:misck_02}. 
The last part of the proof follows from the explicit computation 
$$
(B_\lambda\psi)(x)\;=\;(B\psi)\left(x+\frac{1}{\lambda}\right)\;=\;\int_\R\dd y\; \bb{B}\left(x+\frac{1}{\lambda},y\right)\psi(y)
$$
which provides equation  \eqref{eq:int_ker_B_00}.
\qed

\begin{remark}\label{rk:prescr_B}
{\upshape
Lemma \ref{lemma:B_tild} states that $B_\lambda$ can be expressed as an integral operator only on the dense domain 
$\psi \in L^2(\R)\cap L^1(\R)$. For function in $\psi\in L^2(\R)\setminus L^1(\R)$ in principle, we do not have the right to write $B_\lambda\psi$ using the integral kernel. However, in the following,
 we will tacitly use the following convention
$$
(B_\lambda\psi)(x)\;\equiv\;\lim_{R\to\infty}\int_{-R}^{+R}\dd y\;\bb{B}\left(x+\frac{1}{\lambda},y\right)\psi(y)\;,\quad\text{if}\;\psi\in L^2(\R)\setminus L^1(\R)\;.
$$
This identification must be understood 
as follows: (i) The product $\psi_R:=\psi\chi_{[-R,+R]}$ is in $L^2(\R)\cap L^1(\R)$ 
and so $B_\lambda\psi_R$ can be computed (pointwise) through the integral formula; (ii) $\psi_R\to\psi$,  and in turn $B_\lambda\psi_R\to B_\lambda\psi$,
in the $L^2$-topology; (iii) Then, the identification above  makes sense almost everywhere on subsequences \cite[Theorem 4.9 (a)]{brezis-87}.
}
   \hfill $\blacktriangleleft$
\end{remark}

\medskip

Lemma \ref{lemma:B_tild} allows to describe the domain of $H_T$ as follows:
$$
\s{D}(H_T)\;=\;\left\{\psi\in L^2(\R)\ \big|\ \psi(x)=\int_\R\dd y\; \bb{B}_\lambda(x,y)\phi(y)\;,\quad \phi\in\s{Q}(\R) \right\}\;.
$$
An explicit computation (made of several  changes of integration variable) shows that the generic element $\psi$ in $\s{D}(H_T)$ has the form
$$
\psi(x)\;=\;\frac{1}{x+\frac{1}{\lambda}}\int_0^{+\infty}\dd s \; J_0(\sqrt{s})\;\phi\left(\frac{s}{x+\frac{1}{\lambda}}\right)\;,\quad \phi\in\s{Q}(\R)\;.
$$

\medskip

From \eqref{eq:intro_15} and Theorem \ref{theo:self_ext1D} one infers that 
$\s{S}(\R)\subset \s{D}_0\subset \s{D}(\Pi_0)$ and $\s{S}(\R) +\C[\zeta_0]$ is a core for $\Pi_0$. Since $(\bb{F}S_\lambda)^*[\s{S}(\R)]=\s{S}(\R)$ in view of the invariance of the Schwartz space under the Fourier transform and the translations, it follows that 
$$
\s{D}_0(H_{T})\;:=\;\s{S}(\R)\;+\;\C[\kappa_0]
$$
is a core for $H_{T}$, with
$\kappa_0:=(B_\lambda\bb{F}^*)\eta_0$ (the function $\eta_0$ is described in Proposition \ref{prob:first_step_wp}). Moreover,  the unitary transform $B_\lambda\bb{F}^*$ and Proposition \ref{prob:first_step_wp} also justify \eqref{eq:act_HT_01} with $\kappa_1:=(B_\lambda\bb{F}^*)\eta_\pi$.
\begin{proposition}\label{prop:spec_fun_xi}
The functions $\kappa_0$ and $\kappa_1$ are given by the formulas \eqref{eq:act_HT_02}.
\end{proposition}
\proof
Let $\eta_0(x)=\expo{-|x|}$ and $\eta_\pi(x)=\ii{\rm sgn}(x)\expo{-|x|}$.
The inverse Fourier transforms of these functions are given by
$$
(\bb{F}^*\eta_0)(x)\;=\;\sqrt{\frac{2}{\pi}}\frac{1}{1+x^2}\;,\quad (\bb{F}^*\eta_\pi)(x)\;=\;-\sqrt{\frac{2}{\pi}}\frac{x}{1+x^2}\;.
$$
Since  $\bb{F}^*\eta_0\in L^2(\R)\cap L^1(\R)$, the transformed function $B\bb{F}^*\eta_0$ can be computed via the  integral kernel of $B$. Then
Lemma \ref{lemma:spec_int_funct_appB} provides
$$
\begin{aligned}
(B\bb{F}^*\eta_0)(x)\;&=\;-\ii\sqrt{\frac{8}{\pi}}\;{\rm sgn}(x)\;{\rm kei}\left(2\sqrt{|x|} \right)\;.
\end{aligned}
$$
Since $\bb{F}^*\eta_1\in L^2(\R)\setminus L^1(\R)$, the transformed function $B\bb{F}^*\eta_1$ as to be computed according to the prescription of Remark \ref{rk:prescr_B}. In this case one has
$$
\begin{aligned}
(B\bb{F}^*\eta_\pi)(x)\;&=\;-\sqrt{\frac{2}{\pi}}\lim_{R\to+\infty}\int_{-R}^{+R}\dd y\;  \frac{\bb{B}(x,y)\;y}{1+y^2}\;.
\end{aligned}
$$
However, as shown in the proof of Lemma \ref{lemma:spec_int_funct_appB}, the integrant is absolutely integrable for every values of $x$. This allows  to forget the limit and one gets
$$
\begin{aligned}
(B\bb{F}^*\eta_\pi)(x)\;&=\;\ii\sqrt{\frac{8}{\pi}}\;{\rm ker}\left(2\sqrt{|x|} \right)\;.\\
\end{aligned}
$$
Finally a translation by $S^*_\lambda$ and a multiplication by $-\ii$ provide the formulas \eqref{eq:act_HT_02}.
\qed

\begin{remark}[Other self-joined extensions]\label{rk:more_self_adj-ext}
{\upshape
As for the operator $\Pi$ discussed in Section \ref{sect:aux_op_1D}, also the thermal Hamiltonian $H_T$ admits a family of unitarily equivalent self-adjoint extension parametrized by $\theta\in\n{S}^1$, and defined by
$$
H_{T,\theta}\;:=\;\lambda\; (\bb{F}S_\lambda)^*\Pi_\theta(\bb{F}S_\lambda)\;.
$$
Since $\Pi_\theta=L_\theta\Pi_0L_\theta^*$ one obtains that $H_{T,\theta}$ is related to the standard thermal Hamiltonian $H_T$ by the unitary equivalence
$$
H_{T,\theta}\;:=\;N_\theta H_TN_\theta^*
$$
where $N_\theta:=(\bb{F}S_\lambda)^*L_\theta(\bb{F}S_\lambda)$. An explicit computation provides that
$$
N_\theta\;:=\;\cos\left(\frac{\theta}{2}\right)\;{\bf 1}\;-\; \sin\left(\frac{\theta}{2}\right)\;\rr{H}
$$
where $\rr{H}$ denotes the \emph{Hilbert transform} defined by
$$
(\rr{H}\psi)(x)\;:=\;\frac{1}{\pi}\int_\R\dd y\;\frac{\psi(y)}{x-y}
$$
over sufficiently regular functions $\psi$, and with the integral taken as a Cauchy principal
value.
}
   \hfill $\blacktriangleleft$
\end{remark}

%%%%
\subsection{Unitary propagator}\label{sect:TH_op_1D-case_up}
Let 
$$
U_T(t)\;:=\;\expo{-\ii t H_T}
$$
the \emph{unitary propagator} associated with the self-adjoint operator $H_T$. Using the various unitary equivalences that connect $H_T$ with the momentum operator $p$
one has that 
$$
U_T(t)\;=\;{B}_\lambda\left(\bb{F}^*\expo{-\ii \lambda t p}\bb{F}\right){B}^*_\lambda\;=\;S_\lambda^*({B}\expo{\ii \lambda t x}{B})S_\lambda
$$
where in the last equality we used 
$\bb{F}^*p\bb{F}=-x$. With the help of Lemma \ref{lemma:B_tild} we can compute the integral kernel of $U_T(t)$.
\begin{proposition}\label{lemma:U_propag}
On the dense domain $L^2(\R)\cap L^1(\R)$
the unitary propagator  $U_T(t)$ with ($t\neq 0$) acts as an integral operator with kernel given by \eqref{eq:int_ker_intro_UT} and \eqref{eq:int_ker_intro_UT2}.
\end{proposition}
\proof
Let us start by computing the kernel of
$A_\tau\;:=B\expo{\ii \tau x}{B}$, with $\tau\in\R\setminus \{0\}$, on $\psi\in L^2(\R)\cap L^1(\R)$.
$$
\begin{aligned}
(A_\tau\psi)(x)\;:=\;\lim_{R\to+\infty}\int_{-R}^{+R}\dd y\;\expo{\ii \tau y}\bb{B}(x,y)\left(\int_\R\dd s\;\bb{B}(y,s)\psi(s)\right)
\end{aligned}
$$
The integral in the variable $y$ is meant in the sense of a principal value in view of Remark \ref{rk:prescr_B}. For every $x,\tau\in\R$ the function $g_{(x,\tau)}(s,y):=\expo{\ii \tau y}\bb{B}(x,y)\bb{B}(y,s)\psi(s)$ is absolutely integrable in $\R\times[-R,+R]$ since $|g_{(x,\tau)}|\leqslant  |\bb{B}(x,y)||\psi(s)|$.
Then, we can invoke the 
Fubini-Tonelli theorem to change the order of integration
\begin{equation}\label{eq:maski_002}
(A_\tau\psi)(x)\;=\;\lim_{R\to+\infty}
\int_\R\dd s\;\bb{A}^R_\tau(x,s)\;\psi(s)
\end{equation}
where
$$
\bb{A}^R_\tau(x,s)\;:=\;\int_{-R}^{+R}\dd y\;\expo{\ii \tau y}\bb{B}(x,y)\bb{B}(y,s)\;.
$$
For $xs\neq0$  the change of variables $u:=-xy$ provides
$$
\bb{A}^R_\tau(x,s)\;:=\;\frac{{\rm sgn}(s)+{\rm sgn}(x)}{2x}\int_{0}^{+R|x|}\dd u\;\expo{-\ii \frac{\tau}{x} u}J_0(2\sqrt{u})J_0\left(2\sqrt{\frac{|s|}{|x|}}\sqrt{u}\right)\;.
$$
By using formula \cite[eq. 6.615]{gradshteyn-ryzhik-07}
	one gets
	$$
	\lim_{R\to+\infty}\bb{A}^R_\tau(x,s)\;=\;-\ii\frac{{\rm sgn}(s)+{\rm sgn}(x)}{2 \tau}\expo{\ii\frac{x+s}{\tau}}I_0\left(-\ii\frac{2}{\tau}\sqrt{|xs|}\right)\;.
	$$
Finally, the well known relations $I_0(\pm\ii x)=J_0(\mp x)=J_0(x)$ valid for $x\geqslant 0$ provide
\begin{equation}\label{eq:comp_ker_UT_01}
\lim_{R\to+\infty}\bb{A}^R_\tau(x,s)\;=\;\bb{U}_\tau(x,s)
\end{equation}
where the kernel $\bb{U}_\tau$ is defined by
\eqref{eq:int_ker_intro_UT2}. Equation \eqref{eq:comp_ker_UT_01} is valid also in the singular cases $xs=0$. For instance, for $x=0$ on gets after the usual change of coordinates
$$
\lim_{R\to+\infty}\bb{A}^R_\tau(0,s)\;=\;\frac{1}{2s}\int_0^{+\infty}\dd u\;\expo{-\ii\frac{\tau}{s}u}J_0\left(2\sqrt{u}\right)\;=\;\bb{U}_\tau(0,s)
$$
where the last equality is justified by \cite[eq. 6.614 (1)]{gradshteyn-ryzhik-07}.
The case $s=0$ is similar. In view of \eqref{eq:comp_ker_UT_01} we have the pointwise convergence
$$
\lim_{R\to+\infty}\bb{A}^R_\tau(x,s)\psi(s)\;=\;\bb{U}_\tau(x,s)\psi(s)\;.
$$
and since $|\bb{U}_\tau(x,s)|\leqslant |\tau|^{-1}$ for all $(x,s)\in\R^2$ one has that the function
$s\mapsto\bb{A}^R_\tau(x,s)\psi(s)$ is definitively dominated by the integrable function $s\mapsto|\tau|^{-1}\psi(s)$ (provided $\tau\neq0$). This fact allows  to use the Lebesgue's dominated convergence theorem in \eqref{eq:maski_002}, providing in this way
\begin{equation}\label{eq:maski_003}
(A_\tau\psi)(x)\;=\;
\int_\R\dd s\;\bb{U}_\tau(x,s)\;\psi(s)\;.
\end{equation}
Formula \eqref{eq:int_ker_intro_UT} is obtained by observing that $U_T(t)=S^*_\lambda A_{\lambda t}S_\lambda$.
\qed
\subsection{Resolvent and Green function}\label{sect:TH_op_1D-case_res}
The resolvent of $H_T$ can be computed as the  Laplace transform of the unitary propagator
$U_T(t)$ according to the well known formula \cite[eq. (1.28), p. 484]{kato-95}.
For every $\zeta\in\C\setminus\R$ let
$$
R_\zeta(H_T)\;:=\;\left(H_T-\zeta{\bf 1}\right)^{-1}
$$
be the resolvent of $H_T$. Then, it holds true that
\begin{equation}\label{eq:lapl_res_01}
R_\zeta(H_T)\;=\;\ii\int_0^{+\infty}\dd t\;\expo{\ii\zeta t}\; U_T(t)\;,\qquad{\rm Im}(\zeta)>0
\end{equation}
where the integral is interpreted as a strong Riemann integral $\lim_{\sigma\to+\infty}\int_0^{\sigma}$. The resolvent for ${\rm Im}(\zeta)<0$ can be obtained from the relation
$R_{\overline{\zeta}}(H_T)=R_\zeta(H_T)^*$. The formula \eqref{eq:lapl_res_01} is helpful to compute the integral kernel of $R_\zeta(H_T)$.

\medskip

One can take advantage of the unitary equivalence $U_T(t)=S^*_\lambda A_{\lambda t}S_\lambda$ used in Proposition \ref{lemma:U_propag} to obtain
$R_\zeta(H_T)=\lambda^{-1}S^*_\lambda Z_{\frac{\zeta}{\lambda}} S_\lambda$ with
$$
Z_\alpha\;:=\;\ii\int_0^{+\infty}\dd \tau\;\expo{\ii\alpha \tau}\; A_{\tau}\;,\qquad{\rm Im}(\alpha)>0\;.
$$
Let  $\psi\in L^2(\R)\cap L^1(\R)$. With the integral kernel of $A_{\tau}$ provided in \eqref{eq:maski_003} one can write
$$
(Z_\alpha\psi)(x)\;:=\;\ii\lim_{\sigma\to+\infty}\int_0^{\sigma}\dd \tau\;\expo{\ii\alpha \tau}\left(\int_\R\dd s\;\bb{U}_\tau(x,s)\;\psi(s)\right)\;.
$$
Since the $J_0(\tau^{-1})\sim\sqrt{\tau}$ if $\tau\to 0$ one can check that the function $h_x(\tau,s):=\expo{\ii\alpha \tau}\bb{U}_\tau(x,s)\psi(s)$ 
meets the conditions of the  
Fubini-Tonelli theorem for the change of  the order of integration. Moreover, one can take care of the limit in $\sigma$ with the help of the Lebesgue's dominated convergence theorem. At the end of these manipulations one gets
$$
\begin{aligned}
(Z_\alpha\psi)(x)\;:=\;\int_\R\dd s\;\bb{Z}_\alpha(x,s)\;\psi(s)
\end{aligned}
$$
with kernel
\begin{equation}\label{eq:impl_int_ker_res}
\begin{aligned}
\bb{Z}_\alpha(x,s)\;:&=\;\ii\int_0^{+\infty}\dd\tau\;\expo{\ii\alpha \tau}\bb{U}_\tau(x,s)\\
&=\;\big({\rm sgn}(x)+{\rm sgn}(y)\big)\; F_\alpha(x,y)
\end{aligned}
\end{equation}
where
$$
F_\alpha(x,y)\;:=\;\frac{1}{2}\int_0^{+\infty}\dd\tau\;\frac{\expo{\ii\left(\alpha \tau+\frac{(x+s)}{\tau}\right)}}{\tau}\;J_0\left(\frac{2}{\tau}\sqrt{\left|xs\right|}\right)\;.
$$
Setting $\alpha:=|\alpha|\expo{\ii \phi}$, $0<\phi<\pi$, the last integral 
can be integrated case by case using the
Macdonald's and Nicholson's formulas  \cite[Section 7.7.6]{erdelyi-53} or \cite[Section III, p. 98]{magnus-oberhettinger-66}. A different way of calculating the kernel \eqref{eq:impl_int_ker_res} is sketched
at the end of  Appendix \ref{sect:hank_tras}.
In both cases, after some tedious calculations, one gets
\begin{equation}\label{eq:GOX_03}
\begin{aligned}
F_\alpha(x,y)\;:=\;&I_0\left(2\sqrt{|\alpha|\min\{|x|,|y|\}}\expo{\ii\left[\frac{\phi}{2}-\frac{\pi}{4}\big({\rm sgn}(x)+1\big)\right]}\right)\\
&\;\times\;K_0\left(2\sqrt{|\alpha|\min\{|x|,|y|\}}\expo{\ii\left[\frac{\phi}{2}-\frac{\pi}{4}\big({\rm sgn}(x)+1\big)\right]}\right)\;.
\end{aligned}
\end{equation}
It is also possible to check directly that the kernel $\bb{Z}_\alpha(x,s)$ inverts in a distributional sense the operator $T-\alpha{\bf 1}$.

%--------------%
\subsection{Scattering by a convolution potential}\label{sect:scat_conv_pot}
Let $g\in L^1(\R)$ and consider the associated convolution potential $W_g$ defined by \eqref{eq:scat_int_01}. 
Since $W_g$ is a bounded operator of norm $\|W_g\|=\|g\|_1$ the perturbed operator
$H_{T,g}:=H_T+W_g$ is well defined as a self-adjoint operator on the domain $\s{D}(H_T)$ as a consequence of the Kato-Rellich theorem \cite[Theorem X.12]{reed-simon-II}. The it makes sense to consider the scattering theory of the pair $(H_{T},H_{T,g})$. 

\medskip

In view of the unitary equivalence 
$p=\frac{1}{\lambda}I\bb{F}S_\lambda H_T S^*_\lambda\bb{F}^*I$ between the momentum operator and $H_T$ we can equivalently study the scattering theory of the pair $(p,p_g)$.
where $p_g:=p+M_g$ is the perturbation of the momentum given by the potential
$$
M_g\;:=\:\frac{1}{\lambda}I\bb{F}S_\lambda W_gS^*_\lambda\bb{F}^*I\;.
$$

\begin{lemma}
The potential $M_g$ is the multiplication operator defined by
$$
(M_g\psi)(x)\;:=\;\frac{\sqrt{2\pi}}{\lambda}\hat{g}\left(\frac{1}{x}\right)\psi(x)\;,\qquad \psi\in\L^2(\R)\;.
$$
where $\hat{g}$ denotes the Fourier transform
of $g$.
\end{lemma}
\proof
By construction the convolution is invariant under translations. This means that $S_\lambda W_gS^*_\lambda=W_g$. Moreover the Fourier transform of a convolution gives a multiplication operator
$$
(\bb{F}  W_g\bb{F}^*\psi)(x)\;:=\;\sqrt{2\pi}\hat{g}(x)\psi(x)\;,\qquad\psi\in\L^2(\R)
$$
where $\hat{g}$ denotes the Fourier transform
of $g$. The proof is completed by observing recalling the definition of the involution $I$.
\qed

\medskip

We are now in position to provide the proof of Theorem \ref{theo:main2}.

\proof[Proof (of Theorem \ref{theo:main2})]
Since $g\in L^1(\R)$ then $\hat{g}\in\C_0(\R)$ (continuous functions vanishing at infinity) in view of the Riemann-Lebesgue Lemma \cite[Theorem IX.7]{reed-simon-II}. This implies that the function $x\mapsto \hat{g}\left(\frac{1}{x}\right)$ belongs to $C(\R)\cap L^{\infty}(\R)$. As a result the multiplicative potential $M_g$ is bounded with norm $\|M_g\|=\frac{\sqrt{2\pi}}{\lambda}\|\hat{g}\|_\infty$ and the  conditions of \cite[Example 3.1, p. 530]{kato-95} are satisfied. Then, one obtain that $p$ and $p_g$ are unitarily equivalent. 
This also implies the unitary equivalence of $H_T$ and $H_{T,g}$, and in turn item (i) of claim. In \cite[Example 3.1, p. 530]{kato-95} it is also proven the existence and the completeness for the waves operators associated to the pair $(p,p_g)$ under the assumption that of the existence of the the improper integrals
$$
\begin{aligned}
\lim_{x\to0^+}\int_x^{+\infty}\dd s\; \hat{g}\left(\frac{1}{s}\right)\;&=\;\lim_{x\to+\infty}\int_0^{x}\dd s\; \frac{\hat{g}\left(s\right)}{s^2}\\
\lim_{x\to0^-}\int^x_{-\infty}\dd s\; \hat{g}\left(\frac{1}{s}\right)\;&=\;\lim_{x\to-\infty}\int^0_{x}\dd s\; \frac{\hat{g}\left(s\right)}{s^2}\;.
\end{aligned}
$$
This requires that $\hat{g}\to 0$ fast enough
when $s\to 0^\pm$. This 
is guaranteed by the (not optimal) conditions required in the theorem statement. Invoking once again the unitary equivalence between $p$ and $H_T$ one obtain the existence and the completeness for the waves operators associated to the pair  $(H_T,H_{T,g})$, proving in this way item (ii). 
Also for item (iii), in  \cite[Example 3.1, p. 530]{kato-95} is proven that the $S$-matrix for the pair $(p,p_g)$ is a complex number given by
$$
S_g\;:=\;\expo{-\ii\frac{\sqrt{2\pi}}{\lambda}\int_\R\dd x\; \hat{g}\left(\frac{1}{x}\right)}
\;:=\;\expo{-\ii\frac{\sqrt{2\pi}}{\lambda}\int_\R\dd s\; \frac{\hat{g}\left(s\right)}{s^2}
}\;. 
$$
Since a complex number is unchanged by unitary equivalences it follows that $S_g$ is also the 
$S$-matrix for the pair $(H_T,H_{T,g})$.
\qed

\section{The  classical dynamics}
\label{sect:App_classic}
In this last section  we will study  the classical dynamics induced by a thermal gradient. The classic analogue of the Luttinger's model is provided by the the Hamiltonian function
\begin{equation}\label{eq_HamT1}
H_T(x,p)\;:=\;(1+\lambda\;\gamma\cdot x)\; \frac{p^2}{2m}\;=\;K(p)\;+\;\lambda\; T_\gamma (x,p)\;,\qquad \
\end{equation}
with parameters $\lambda>0$ and $\gamma\in\n{S}^{d-1}$. The Hamiltonian $H_T$
can be seen as the sum of the Hamiltonian of a \emph{free} $d$-dimensional particle of mass $m$
$$
K(p)\;:=\;\frac{p^2}{2m}\;=\;\frac{1}{2m}\sum_{j=1}^Np_j^2
$$
coupled through the \emph{coupling constant} $\lambda>0$ with the \emph{thermal potential} 
$$
T_\gamma(x,p)\;:=\;(\gamma\cdot x)\;K(p)\;=\;\left(\frac{p^2}{2m}\right)\;\sum_{j=1}^d\gamma_j\; x_j
$$
along the direction $\gamma\in\n{S}^{d-1}$. The coupling constant has the dimension of the inverse of a distance, namely 
$
\lambda={\ell^{-1}}
$
 with $\ell>0$  the  \emph{typical length} of the thermal field. Therefore, the limit $\lambda\to 0$ 
 describes  the situation in which the  typical length of the field is  much larger than the  typical length of the system (\eg the size of the particle).
The potential $T_\gamma$ is an example of  what is known as a \emph{generalized potential}, namely a potential which depends not only on the position but also on the  the velocity.

%%%%%%%%%%%%%
\subsection{Hamiltonian Formalism and Newton equation}
The Hamilton equations associated to \eqref{eq_HamT1} read
\begin{equation}\label{eq:ham_sys0001}
\left\{
\begin{aligned}
\dot{x}\;&=+{\nabla}_{p}  H_T\;=\;\frac{\left(1+\lambda\;{\gamma}\cdot{x} \right)}{m}\;\boldsymbol{p}\\
\dot{p}\;&=-{\nabla}_{x}  H_T\;=\;-\lambda\;\frac{p^2}{2m}\;\gamma\;.
\end{aligned}
\right.
\end{equation}
The first equation  can be inverted  out of the \emph{critical plane} 
\begin{equation}\label{eq:crit_plane}
\Xi_{\rm c}\;:=\;\big\{x\in\R^{d}\ |\ \gamma\cdot{x}+\ell=0
\big\}
\end{equation}
and provides
\begin{equation}\label{eq:mom-vel}
{p}({x},\dot{x})\;=\;\frac{m}{\left(1+\lambda\;{\gamma}\cdot{x} \right)}\;\dot{x}
\end{equation}
One can restore the usual relation  ${p}=m_T \dot{x}$ between  momentum and velocity by introducing the \emph{position-dependent mass} (PDM)
$$
m_T(x)\;:=\;\frac{m}{\left(1+\lambda\;{\gamma}\cdot{x} \right)}\;.
$$
It is interesting to notice that the Hamiltonian \eqref{eq_HamT1} can be rewritten as
\begin{equation}\label{eq_HamT2}
H_T(x,p)\;=\; \frac{p^2}{2m_T(x)}\;,
\end{equation}
namely as the Hamiltonian of a free particle with a PDM. The second equation of \eqref{eq:ham_sys0001} can be rewritten as
\begin{equation}\label{eq:p_dot_Theta}
\dot{p}\;=\;-\lambda\;{\nabla}_{x}T_\gamma \;.\end{equation}

\medskip

A straightforward computation allows to derive the Newton's laws from \eqref{eq:ham_sys0001}:
$$
\begin{aligned}
m\;\ddot{x}\;&=\; \lambda(\gamma\cdot\dot{x}){p}\;-\;\lambda\left(1+\lambda\;{\gamma}\cdot{x} \right)\frac{p^2}{2m}\;\gamma\;.
\end{aligned}
$$
After introducing \eqref{eq:mom-vel} in the las expression one obtains the Newton's equation
$$
m\;\ddot{x}\;=\; \lambda\;F_T(x,\dot{x})
$$
where the \emph{thermal force} (which has the   dimensions of a force times a distance) is given by
\begin{equation}\label{eq_tern-forc}
{F}_T(x,\dot{x})\;=\; m_T(x)\;\left[(
\gamma \cdot\dot{x}
)\dot{x}\;-\;\frac{\dot{x}^2}{2}\;\gamma\right].
\end{equation}
A way of interpreting this  Newton's Equation is to say that  the motion of the PDM-particle is   influenced by the effect of its
own internally self-produced  force field generated by the spatial dependence of the mass.
The relation between the force $F_T$ and the potential $T_\gamma$ can be deduced by observing that
\begin{equation}\label{eq:p_dot}
-\lambda{\nabla}_{x}T_{\gamma}\;=\; -\lambda\;\frac{m_T(x)^2}{m}\frac{\dot{x}^2}{2}\;\gamma\;
\end{equation}
in view of the \eqref{eq:p_dot_Theta}, \eqref{eq:ham_sys0001} and \eqref{eq:mom-vel}, respectively.
After some manipulation and the use of equation \eqref{eq:mom-vel} one gets
\begin{equation}\label{eq_tern-forc2}
\begin{aligned}
{F}_T(x,p)\;&=\; 
- {\nabla}_{x}T_\gamma(x,p)\;+\;{R}_T(x,p)
\end{aligned}
\end{equation}
which shows that the thermal force is not simply given by $-{\nabla}_{x}T_\gamma$,  as for ordinary conservative forces, but it includes an extra \emph{reacting} term 
\begin{equation}\label{eq:reac_term}
R_T(x,p)\;:=\;\frac{\dd}{\dd t} \left(
\gamma \cdot{x}
\right)p\;=\; m\;\frac{\dd}{\dd t} \left({\nabla}_{p}T_\gamma(x,p)
\right)\;
\end{equation}
which is generally not aligned with the direction $\gamma$
of the field.

%%%%%%%%%%%%%
\subsection{Qualitative analysis}\label{sec:qual_analys}
Let us start with the analysis of the qualitative behavior of the solution of the Hamiltonian system \eqref{eq_HamT1}. To simplify the study let us fix  convenient notations. The unit vector $\gamma$ can be completed to an orthonormal basis by adding other $d-1$  orthonormal vectors $e_1,\ldots,{e}_{d-1}$. This allows to fix the generalized coordinates $x_0:=\gamma\cdot x$, $x_j:= e_j\cdot x$, and the generalized momenta $p_0:=\gamma\cdot p$, $p_j:=e_j\cdot p$ with $j=1,\ldots,d-1$.
 In this coordinates  the  Hamiltonian \eqref{eq_HamT1}  reads
\begin{equation}\label{eq:00XX01}
H_T(x_0,p_1,\ldots,p_d)\;=\;\left(1+\lambda {x_0} \right)\; \frac{p^2}{2m}\;
\end{equation}
and the  Hamilton equations \eqref{eq:ham_sys0001} become
\begin{equation}\label{eq:ham_sys01}
\left\{
\begin{aligned}
\dot{x}_j\;&=\;\left(1+\lambda\;{x_0}\right)\; \frac{p_j}{m}\\
\dot{p}_j\;&=\;-  \delta_{0,j}\;\lambda\;\frac{p^2}{2m}
\end{aligned}
\qquad\quad j=0,\ldots,d-1\;.
\right.
\end{equation}
The integration of the equations for the \virg{orthogonal}   components of the momentum  immediately leads to
$$
p_j(t)\;=\;\wp_{j}\;=\;{\rm const.}\,,\qquad\quad j=1,\ldots,d-1\;.
$$
This can   be seen as a consequence of the Noether's theorem applied to the invariance under translations of the Hamiltonian $H_T$ along all the  directions  orthogonal to $\gamma$.
Let us introduce the constant of motion 
$$
\wp_\bot:=\left(\sum_{j=1}^{d-1}\wp_{j}^2\right)^{\frac{1}{2}}
$$
which quantifies the momentum in   the orthogonal  plane to the direction of the  thermal field. The square   of the momentum at any time takes the form
\begin{equation}\label{eq:tot_moment_t}
{p}^2(t)\;=\;p_0^2(t)\;+\;\wp_\bot^2\;.
\end{equation}

\medskip

The value of the parameter $\wp_\bot$ strongly determines the behavior   of the solutions of the system \eqref{eq:ham_sys01}. To see this, one can observe  that  the Hamiltonian $H_T$ is  time-independent
and therefore  the Noether's theorem provides a further constant of motion, \ie the (total) energy
$$
E_0\;:=\;\left(1+\lambda\;\varrho_0\right)\; \frac{\wp_0^2+ \wp_\bot^2}{2m}
$$ 
which is completely specified by the initial  conditions 
$$
\varrho_0\;:=\;x_0(t=0)\;,\qquad\qquad \wp_0\;:=\;p_0(t=0)\;.
$$
 The constraint
\begin{equation}\label{eq:time_const_ener}
H_T(x(t),p(t))\;=\;E_0\;,\qquad\quad\forall\ t\in\R
\end{equation}
can be used to obtain the equation 
\begin{equation}\label{eq:time_X1}
x_0(t)\;=\;\frac{1}{\lambda}\left(\frac{2m E_0}{p^2(t)}\;-\; 1\right)\;=\;\frac{\wp_0^2+ \wp_\bot^2}{p_0^2(t)+\wp_\bot^2}\left(\frac{1}{\lambda}\;+\;\varrho_0\right)\;-\; \frac{1}{\lambda}\;,
\end{equation}
which provides the time evolution of $x_0$ once it is known the form of $p_0^2(t)$ and the initial conditions $\varrho_0$ and $\wp_0,\wp_1,\ldots,\wp_{N-1}$. In addition to this, the constraint \eqref{eq:time_const_ener} also provides   useful information for a qualitative study of the trajectory $x(t)$ of the particle. A comparison between \eqref{eq:00XX01} and \eqref{eq:time_const_ener} shows that the sign of  $E_0$ only depends on the quantity $1+\lambda \varrho_0$. More precisely,   one has that
$$
\begin{aligned}
\pm E_0\;\geqslant\;0&\qquad \Leftrightarrow \qquad&  \pm\varrho_0\;\geqslant\;\mp\ell\;\;.\\
\end{aligned}
$$
Thus, the  {critical plane} $\Xi_{\rm c}\subset\R^N$   separates the space into two regions labelled by the sign of the energy $E_0$. The full trajectory  $x(t)$ of the particle is fully contained in only one of these two half-spaces
according to the initial position $\varrho_0$ along the direction $\gamma$ at the initial time $t=0$. Moreover, the trajectory can touch the critical plane only 
 at the cost of a divergence in the value of the total momentum, $p^2\to\infty$.

\medskip

The existence of this critical impenetrable plane can be justified on the basis of the Newton's law $
m \ddot{x}_j=\lambda\;F_{T,j}\;
$
where the force \eqref{eq_tern-forc} is given for components by
\begin{equation}\label{eq_tern-forc2.0}
F_{T,j}\;=\;
\left\{
\begin{aligned}
& \frac{E_0}{2}\;-\;\left(1+\lambda {x_{0}}\right)
\frac{\wp_\bot^2}{m}\
 &\qquad \text{if}&\quad j=0\\
&\left(1+\lambda{x_{0}}\right)\; 
\frac{{p}_{0}{\wp}_{j}}{m}&\qquad \text{if}&\quad j=1,\ldots,d-1\;.\\
\end{aligned}
\right.
\end{equation}
In the derivation of   \eqref{eq_tern-forc2.0} from  \eqref{eq_tern-forc} we  made use of  \eqref{eq_HamT2} along with $m_T\dot{x}=p$ and the conservation laws \eqref{eq:tot_moment_t} and \eqref{eq:time_const_ener}. The component $F_{T,0}$ is proportional to $E_0$ very close to the critical plane ($1+\lambda {x_{0}}\sim0$) and force the  particle to stay inside the half-space where the particle was at the initial time. 
When $\wp_\bot^2\neq0$ the component $F_{T,0}$ changes sign sufficiently far from the critical plane and begins to attract the particle towards $\Xi_{\rm c}$. This suggests that the motion of the particle must be bounded in the direction $\gamma$ provided that the momentum has a non-vanishing component 
orthogonal to $\gamma$ at the initial time.
 The components $F_{T,1},\ldots,F_{T,d-1}$ are due to the reaction term $R_T$ \eqref{eq:reac_term}. The conservation of the energy implies that $|p_0|\propto |1+\lambda {x_{0}}|^{-\frac{1}{2}}$ for $x_0\to -\ell$. Therefore the  orthogonal components of  $F_T$ vanish when the particle approaches the critical plane.

%%%%%%%%%%%%%
\subsection{Exceptional solutions}
The Hamilton equations \eqref{eq:ham_sys01} (or equivalently \eqref{eq:ham_sys0001}) admit the \emph{exceptional} family of solutions $p(t)=0$ and $x(t)={\varrho}$ for all $t\in\R$ parametrized by all the possible initial positions  $\varrho\in\R^{d}\setminus \Xi_{\rm c}$ 
not belong to the critical plane.
In this case
the  particle is at every moment
 at  rest in a configuration of  total zero energy $E_0=0$. This is not surprising even though the particle is  immersed in the thermal field. In fact the force $F_T$ produced by the    field
vanishes when  $p=0$. If at the initial time one has 
 $\wp_j=0$ for all $j=0,\ldots,d-1$ and  $\varrho_0\neq-\ell$, then ${p}^2 =0$ for all $t\in\R$ (as a consequence of energy conservation) and  therefore  the particle is not subject to any force. This allows the particle to stay in equilibrium forever at the position $\varrho$. 

\medskip 
 
 Another family of exceptional solutions is again described  by   
 $x(t)={\varrho}$ for all $t\in\R$ with the initial positions ${\varrho}\in \Xi_{\rm c}$.
Also in this case the  particle remains  at rest in a configuration of total zero energy $E_0=0$. However, since the particle lies in the critical plane the total momentum is not forced to be zero. While the component of the momentum  orthogonal to  $\gamma$  is constant  and quantified by $\wp_\bot$ the component $p_0(t)$ evolves in time according to the Hamilton equation \eqref{eq:ham_sys01} (with solutions \eqref{eq:pt-1D} if $\wp_\bot=0$ or \eqref{eq:p_1t} when $\wp_\bot\neq0$).

%%%%%%%%%%%%%
\subsection{The general solution}\label{Sec:gen_one-d-case}  
Let us derive the general solution of the Hamiltonian system \eqref{eq:ham_sys01} under the generic  assumption
$\wp_\bot\neq 0$. In this case the differential equation for $p_0$ reads
$$
\dot{p}_0 \;=\; -\lambda\;\frac{p_0^2\;+\;\wp_\bot^2}{2m}
$$
and is solved by
\begin{equation}\label{eq:p_1t}
{p}_0(t)\;=\;\wp_\bot\;\tan\left( \phi\;-\;\lambda\frac{\wp_\bot}{2m}  t\right)
\end{equation}
where $\phi:={\rm arctan}\left(\frac{\wp_0}{\wp_\bot}\right)$ is determined by the  initial conditions.
Equation \eqref{eq:p_1t} shows that  ${p}_0(t)$ diverges periodically at the critical times 
$t_{\rm c}^{(n)}:=t_{\rm c}+n T$,
  $n\in\Z$,
where 
$$
t_{\rm c}\;:=\; \left(2\phi- {\pi} \right)\frac{\ell m}{\wp_\bot}\;,\qquad\quad T\;:=\;2\pi\frac{\ell m}{\wp_\bot}\;
$$
and $\ell=\lambda^{-1}$.

\medskip

From  \eqref{eq:p_1t} and \eqref{eq:tot_moment_t}
one  immediately gets
$$
p^2(t)\;=\;
\frac{\wp_\bot^2}{\cos\left(\phi- \lambda\frac{\wp_\bot}{2m}t\right)^2}
$$
and after some manipulations, equation \eqref{eq:time_X1} provides 
\begin{equation}\label{eq:x_1t_gen_cas}
x_0(t)\;=\;\varrho_0\;+\;A_\lambda \left[\cos\left(\phi- \lambda\frac{\wp_\bot}{2m}t\right)^2\;-\;\cos(\phi)^2\right] 
\end{equation}
where we    the amplitude $A_\lambda$ is given by
$$
A_\lambda\;:=\;\ell \frac{2m E_0}{\wp_\bot^2}\;=\; \frac{\ell+\varrho_0}{\cos(\phi)^2}\;.
$$
Equation \eqref{eq:x_1t_gen_cas} shows that the motion along the direction $\gamma$ is bounded and more precisely is confined between the 
critical plane $\Xi_{\rm c}$ which is reached periodically at the critical times $t_{\rm c}^{(n)}$ and the \emph{extremal} plane 
\begin{equation}\label{eq:extr_plane}
\Xi_{\rm e}\;:=\;\left\{x\in\R^d\ |\ \gamma\cdot x=\varrho_0+\left(\frac{\wp_0}{\wp_\bot}\right)^2\left(\ell+\varrho_0\right)
\right\}
\end{equation}
which is reached periodically at the \emph{extremal} times $t_{\rm e}^{(n)}:=t_{\rm e}+nT$ where $
t_{\rm e}:= 2\phi\frac{\ell m}{\wp_\bot}$.

\medskip

By inserting the solution \eqref{eq:x_1t_gen_cas} in the
differential equations for the other components of the position one gets
$$
\dot{x}_j(t)\;=\; \lambda\frac{\wp_j}{m} A_\lambda \cos\left( \phi- \lambda\frac{\wp_\bot}{2m}t\right)^2\;,\qquad\quad j=1,\ldots,d-1\;.
$$
For each $j$, the corresponding differential equation is integrated by
\begin{equation}\label{eq:x_jt_gen_cas}
x_j(t)\;=\;\varrho_j\;+\;\lambda \frac{\wp_j}{2m} A_\lambda t\;-\;\frac{A_\lambda}{2}\;\frac{\wp_j}{\wp_\bot}\;\left[ \sin\left( 2\phi- \lambda\frac{\wp_\bot}{m}t\right)\;-\;\sin( 2\phi)\right]\;.
\end{equation}
Evidently the motion in the directions $e_j$ is unbounded when $\wp_j\neq 0$
due to the linear term in $t$ which describes a uniform motion with constant velocity $v_{j,\lambda}:=\lambda A_\lambda \frac{\wp_j }{2m}$.  

\medskip

Let us introduce the unit vector  $\nu:=\wp_\bot^{-1}\sum_{j=1}^{d-1}\wp_j e_j$. By construction $\nu$ is orthogonal  to $\gamma$ and $\wp:=\wp_0 \gamma+ \wp_\bot\nu$
 describes the initial momentum of the particle at $t=0$. From \eqref{eq:x_1t_gen_cas} and \eqref{eq:x_jt_gen_cas} one gets that 
\begin{equation}\label{eq:traject}
x(t)\;=\; \varrho\;+\;A_\lambda\big(f_0(t)\gamma\;+\;f_\bot(t)\nu\big)
\end{equation}
with $\varrho:=\varrho\gamma+\sum_{j=1}^{d-1}\rho_j e_j$ the initial position
and
$$
\begin{aligned}
f_0(t)\;&:=\;\cos\left(\phi- \lambda\frac{\wp_\bot}{2m}t\right)^2\;-\;\cos(\phi)^2 \\
f_\bot(t)\;&:=\;\lambda \frac{\wp_\bot}{2m} t\;-\;\frac{1}{2}\;\left[ \sin\left( 2\phi- \lambda\frac{\wp_\bot}{m}t\right)\;-\;\sin( 2\phi)\right]\;.\\
\end{aligned}
$$
Equation \eqref{eq:traject} shows that the motion of the particle is essentially two-dimensional. In fact the orbit $x(t)$ lies entirely in the affine  plane spanned by $\mu$ and $\nu$ and passing through the initial position $\rho$.

\begin{remark}[2D-case]\label{rk:class_2D}
{\upshape
In view of \eqref{eq:traject} the general motion of a particle in the thermal field is a two-dimensional motion provided that the initial momentum is not aligned with the direction of the field. Therefore, one can always  identify the direction $\gamma$ of the field and the direction $\nu$ of the orthogonal component of the  initial momentum 
with the  $x$-axis  and the  $y$-axis of $\R^2$, respectively. This allows us to use the \virg{cozy} notation $x(t)$ and $y(t)$ for the two projections of the trajectory along the direction $\gamma$ y $\nu$, respectively. Let $\wp=(\wp_x,\wp_y)$ be the components of the initial momentum projected along  the two coordinate direction $\gamma$ and $\nu$.
Let us consider here the special situation in which the total momentum is completely orthogonal to $\gamma$. This means that  $\wp_0=\wp_x=0$ and $\wp_\bot=|\wp_y|=|\wp|$. This also implies that $\phi=\arctan(0)=0$ and $A_\lambda=\ell+\varrho_x$ with $\varrho_x=\varrho_0$ is the $x$-component of the initial position $\varrho=(\varrho_x,\varrho_y)$. In this case the 
equations of motion for the position simplify to 
\begin{equation}\label{eq:position-2D}
\begin{aligned}
x(t)\;&=\;\varrho_x\;+\;(\ell+\varrho_x)\left[\cos\left( \lambda\frac{|\wp|}{2m}t\right)^2\;-\;1\right]\;, \\
\\
y(t)\;&=\;\varrho_y\;+\;(\ell+\varrho_x)\left[\lambda\frac{|\wp|}{2m}t\;+\;\frac{1}{2}\; \sin\left(  \lambda\frac{|\wp|}{m}t\right)\right]\;.\\
\end{aligned}
\end{equation}
The 
time evolution of the momentum is described by the equations ${p}_x(t)=-|\wp|\;\tan\left( \lambda\frac{|\wp|}{2 m}t\right)$ and ${p}_y(t)=\wp_y$.
}
   \hfill $\blacktriangleleft$
\end{remark}

%%%%%%%%%%%%%
\subsection{The one-dimensional case}\label{Sec:one-d-case} 
As discussed at the end of Section \ref{Sec:gen_one-d-case} (see Remark \ref{rk:class_2D})   the general motion of a particle in the thermal field is two-dimensional whenever $\wp_\bot\neq0$. Therefore the condition  $\wp_\bot=0$, $\wp_0\neq 0$ corresponds to considering the one-dimensional case. In fact, under these conditions, one immediately gets from \eqref{eq:ham_sys01} that $p_j(t)=\wp_{ j}=0$
for all $j=1,\ldots,d-1$. This  in turn implies $\dot{x}_j=0$ for $j=1,\ldots,d-1$ and so 
$$
x_j(t)\;:=\;\varrho_j\;=\;{\rm const.}\,,\qquad\quad j=1,\ldots,d-1\;.
$$
This means that the only possible motion could take place exclusively in the direction $\gamma$, namely it is one-dimensional. 

\medskip

Without loss of generality let us assume that $\varrho_1=\ldots=\varrho_{d-1}=0$ which means that $x_j(t)=0=p_j(t)$ for all $j=1,\ldots,d-1$.
Given that, the only interesting degrees of freedom are $x_0$ and $p_0$ and we can simplify the notation 
identifying $x_0$ with $x$ and $p_0$ with $p$. With this notation the (non-trivial) one-dimensional system of  Hamilton equations reads
\begin{equation}\label{eq:ham_sys01_1D}
\left\{
\begin{aligned}
\dot{x}\;&=\;\left(1+\lambda\;{x}\right) \frac{p}{m}\\
\dot{p}\;&=\;-  \lambda\frac{p^2}{2m}\;.
\end{aligned}
\right.
\end{equation}
The equation for the momentum immediately integrated by
\begin{equation}\label{eq:pt-1D}
{p} (t)\;=\;\ell\frac{\wp}{\frac{\wp}{2m}\;t+ \ell}
\end{equation}
with $\wp=p(0)$ the initial momentum. Notice that the value of the momentum diverges at the critical time 
$t_{\rm c}:=-\ell\frac{2m}{\wp}$.

\medskip

The time evolution of the position can be derived directly from equation \eqref{eq:time_X1} which, after some algebraic manipulation, provides  
\begin{equation}\label{eq:xt-1D}
\begin{aligned}
{x}(t)\;
&=\;\frac{\left(\ell+ {\varrho}\right)}{\ell^2}\left(\frac{\wp}{2m}\;t+ \ell\right)^2\;-\;\ell\;
\end{aligned}
\end{equation}
with $\varrho=x(0)$ the initial position.
The long time behavior of the trajectory is  determined by the sign of the coefficient of $t^2$ in \eqref{eq:xt-1D}, namely by the sign of  $\ell+ \varrho$.
It follows that 
$$
\lim_{|t|\to\infty}{x}(t)\;=\;\pm\infty\qquad\quad \text{if}\qquad \pm \varrho\;>\; \mp \ell\;.
$$
The turning time in which the velocity changes
sign is determined by  $\dot{x}(t)=0$ and a simple computation shows that this time coincides with the critical time $t_{\rm c}$. Moreover, one has that  
$x(t_{\rm c})=-\ell$ independently
of the initial value $\varrho_0\neq -\ell$. 
In conclusion the critical plane  $\Xi_{\rm c}$  separates the space into two regions and the trajectory   $x(t)$ is fully contained in only one of these two half-spaces
according to  the initial position $\varrho$. Moreover, the trajectory can touch the critical plane only once at the critical time $t_{\rm c}$. These results are in accordance with the qualitative analysis of Section \ref{sec:qual_analys}.

%%%%%%%%%%%%%
\subsection{The Lagrangian Formalism}\label{sect:lagrangian}
By using the Legendre transformation 
$
\bb{L}_T(x,\dot{x})=\dot{x}\cdot p\;-\;H_T(x,p)
$
one can compute the Lagrangian of the system:
\begin{equation}\label{eq:lagPDM}
\bb{L}_T(x,\dot{x})\;:=\;\frac{1}{2}\frac{m}{\left(1+\lambda\;\gamma\cdot{x} \right)}\;\dot{x}^2\;=\;m_T(x)\;\frac{\dot{x}^2}{2}\;.
\end{equation}
Expressions of the type \eqref{eq:lagPDM} are well studied in the literature under the name of \emph{quasi-free  PDM  Lagrangian} (see \cite{mazharimousavi-mustaf-13,bagchi-das-ghosh-poria-13,mustafa-13} and references therein). 
The canonical momentum 
$$
{p}(x,\dot{x})\;:=\; {\nabla}_{\dot{x}}\bb{L}_T(x,\dot{x})\;=\;m_T({x})\;\dot{x}
$$
is exactly that given by equation \eqref{eq:mom-vel}. To compute the Euler-Lagrange  equations of motion we need also 
$$
{\nabla}_{x}\bb{L}_T(x,\dot{x})\;=\; {\nabla}_{x}m_T(x)\;\frac{\dot{x}^2}{2}\;=\;-\lambda\;\frac{m_T({x})^2}{m}\frac{\dot{{x}}^2}{2}\;\gamma\;.
$$
A comparison with \eqref{eq:p_dot} shows that
$$
{\nabla}_{x}\bb{L}_T\;=\;\dot{p}\;=\;- {\nabla}_{x}  H_T
$$
and this assures that the Euler-Lagrange  equation
$$
\frac{\dd}{\dd t}\left( {\nabla}_{\dot{x}}\bb{L}_T\right)\;=\;{\nabla}_{x}\bb{L}_T
$$
is equivalent to the Hamilton system \eqref{eq:ham_sys0001}. An explicit computation provides 
$$
\begin{aligned}
\frac{\dd}{\dd t}\left( {\nabla}_{\dot{x}}\bb{L}_T\right)\;=\;m_T(x)\;\ddot{x}\;+\;\frac{\dd}{\dd t}\left(m_T({x})\right)\;\dot{{x}}\\
\;=\;m_T({x})\;\ddot{{x}}\;-\;\lambda\;\frac{m_T({x})^2}{m}\; \left(
\gamma \cdot\dot{x}
\right)\dot{{x}}
\end{aligned}
$$
and putting all the pieces together one gets
\begin{equation}\label{eq:lagPDM2A}
m_T(x)\;\ddot{x}\;=\;\lambda\;\frac{m_T(x)^2}{m}\; \left(
\gamma \cdot\dot{x}
\right)\dot{x}\;-\;\lambda\;\frac{m_T({x})^2}{m}\frac{\dot{{x}}^2}{2}\;\gamma
\end{equation}
which is equivalent to the  Newton's equation
$m \ddot{x}= \lambda F_T$ with the force \eqref{eq_tern-forc}.

\medskip

In the one-dimensional it is useful to  use  the  change of Lagrangian coordinates $(x,\dot{x})\mapsto (q,\dot{q})$ implemented by
$$
x(q)\;:=\;\expo{\lambda q}\;-\;\frac{1}{\lambda}\;,\qquad\quad \dot{x}(q,\dot{q})\;:=\;\lambda\;\expo{\lambda q}\;\dot{q}\;.
$$
The inverse  is given by
$$
q(x)\;:=\;\frac{1}{\kappa}\log\left(x+\frac{1}{\lambda}\right)
$$
and shows that the change of  coordinates between $x$ and $q$ is one-to-one only when $x\geqslant-\ell$. However, as seen in Section \ref{sec:qual_analys}, this is exactly the range of values of interest for the problem. With this change of coordinates the Lagrangian becomes
\begin{equation}\label{eq:lagPDM-new_coord}
\bb{L}'_T(q,\dot{q})\;:=\;m\;\lambda\;\expo{\lambda q}\;\frac{\dot{q}^2}{2}\;.
\end{equation}
and the associated  Euler-Lagrange  equation reads
$$
\ddot{q}\;:=\;-\lambda\;\frac{\dot{q}^2}{2}\;.
$$
This equation immediately provides the time-behavior of the generalized velocity
$$
\dot{q}(t)\;:=\;\frac{\dot{q}_0}{1+\frac{\dot{q}_0\lambda}{2}t}
$$
and a further integration gives
$$
{q}(t)\;:=\;q_0\;+\;\frac{2}{\lambda}\log\left(1+\frac{\dot{q}_0\lambda}{2}t\right)
$$
where ${q}_0,\dot{q}_0$ are the initial conditions. By coming back to the original variable one can 
recover the expression \eqref{eq:xt-1D} for  $x(t)$.

\appendix

%--------------------%
%--------------------%
\section{Spectral theory of the momentum operator}
\label{sect:app_spec_p}
Let $p=-\ii\frac{\dd}{\dd x}$ be the momentum operator with domain $H^1(\R)\subset L^2(\R)$.
and purely absolutely continuous spectrum 
$\sigma(p)=\sigma_{\rm a.c.}(p)=\R$.
%
%-----%
\subsection{Green's function}\label{app:B1}
With the help of the Fourier transform $\bb{F}$ one gets \cite[IX.29]{reed-simon-II}
$$
\big(\left(p-\zeta{\bf 1}\right)^{-1}\psi\big)(x)\;=\;\int_{\R}\dd y\; \bb{G}^0_\zeta(x,y)\;\psi(y)
$$
where the Green's function of  $p$ is given by
$$
\bb{G}^0_{\zeta}(x,y)\;:=\;\frac{1}{\sqrt{2\pi}}\;\bb{F}^*\left[\frac{1}{k-\zeta}\right](x-y)\;.
$$
A  straightforward computation involving contours integrals in the complex plane provides
\begin{equation}\label{eq:green_p}
\bb{G}^0_{\epsilon\pm\ii\delta}(x,y)\;=\;\pm\ii  \Theta\big(\pm(x-y)\big)\expo{\ii\epsilon(x-y)}\;\expo{-{\delta}\;|x-y|}\end{equation}
with $\epsilon\in\R$ and $\delta>0$.

\subsection{Spectral measure}
\label{app:B2}
Let $\mu_\psi^A$ be the spectral measure of the self-adjoint operator $A$ associated with the state   $\psi\in L^2(\R)$. The function $F_\psi^A:\C\setminus\R\to\C$ defined by the scalar product
$$
F_\psi^A(\zeta)\;:=\;\langle\psi,(A-\zeta{\bf 1})^{-1} \psi\rangle\;=\;\int_\R\dd \mu_\psi^A(\epsilon)\;\frac{1}{\epsilon-\zeta}
$$
is called the \emph{Borel-Stieltjes transformation} of the finite Borel measures $\mu_\psi^{A}$. Since
$$
{\rm Im}\left(F_\psi^{A}(\zeta)\right)\;=\;{\rm Im}(\zeta)\;\int_\R\dd \mu_\psi^{A}(\epsilon)\;\frac{1}{|\epsilon-\zeta|^2}
$$
it follows that $F_\psi^{A}:\C^+\to\C^+$ is is a holomorphic map from the upper half plane $\C^+$ into itself. Such functions are called \emph{Herglotz} or \emph{Nevanlinna functions} (see \cite[Section 1.4]{demuth-krishna-05} or \cite[Appendix]{aizenman-warzel-15}). 
A classical result by de la Vall\'{e}e-Poussin assures that the limit $F_\psi^{A}(\epsilon):=\lim_{\delta\to 0^+}F_\psi^{A}(\epsilon+\ii\delta)$ exists and is finite for Lebesgue-almost every $\epsilon\in\R$. Moreover,  the absolutely continuous part of the spectral measure $\mu_\psi^{A}$ can be recovered from the imaginary part of  $F_\psi^{A}(\rho)$ according to the classical formula
\cite[Theorem 1.4.16.]{demuth-krishna-05} 
$$
\mu_\psi^{A}|_{\rm a.c.}(\dd \epsilon)\;=\;f^A_\psi(\epsilon)\;\dd\epsilon
$$
with
\begin{equation}\label{eq:dens_mes1}
f^A_\psi(\epsilon)\;:=\;\lim_{\delta\to 0^+}\frac{1}{\pi}\;{\rm Im}\left(F_\psi^{A}(\epsilon+\ii\delta)\right)\;.
\end{equation}

\medskip

In the case $A=p$ is the standard momentum operator one  knows that the spectral measure
 is purely absolutely continuous, \ie  $\mu_\psi^{p}=\mu_\psi^{p}|_{\rm a.c.}$.
By the help fo the Fourier transform $\bb{F}$ one obtains that
$$
\begin{aligned}
&F_\psi^p(\epsilon+\ii\delta)\;:=\;\langle\psi,(p-(\epsilon+\ii\delta){\bf 1})^{-1} \psi\rangle\;=\;\int_\R\dd k\;\frac{|\widehat{\psi}(k)|^2}{(k-\epsilon)-\ii\delta}\\
\end{aligned}
$$
where $\widehat{\psi}:=\bb{F}(\psi)$ is the Fourier transform of $\psi$. The application of the formula \eqref{eq:dens_mes1} provides 
$$
f_\psi^p(\epsilon)\;=\;\lim_{\delta\to 0^+}
\int_\R\dd k\;\frac{1}{\pi}\frac{{\delta}}{(k-\epsilon)^2+\delta^2}|\widehat{\psi}(k)|^2\;=\;
|\widehat{\psi}(\epsilon)|^2
$$
where in the last equality one used that $\frac{1}{\pi}\frac{\delta}{x^2+\delta^2}$
converges in the distributional sense to $\delta(x)$ when ${\delta}\to0^+$. In this way one recovers the well-known result
\begin{equation}\label{eq:spec_meas_p}
\mu_\psi^{p}(\dd\epsilon)\;=\;|\widehat{\psi}(\epsilon)|^2\;\dd\epsilon\;.
\end{equation}
\subsection{Density of states}\label{app:dens_stat_p}
For $\epsilon\in\R$ let $P_\epsilon$, 
be the spectral projection of $p$ associated with the energy $\epsilon$ according to \eqref{eq:density_T_02}. Let
$(Q_L\psi)(x)=\chi_{[-L,L]}(x)\psi(x)$ be  the projection which restricts the functions $\psi\in L^2(\R)$  on the set $[-L,L]$. The quantity
$$
\s{N}^p_L(\epsilon)\;:=\;\frac{{\rm sgn}(\epsilon)}{2L}\;{\rm Tr}\left(P_\epsilon Q_L\right)
$$
is well defined since $P_\epsilon\Q_L$ is trace-class in view of \cite[Theorem XI 20]{reed-simon-III}. 
\begin{lemma}\label{lemma:IDOS_p}
For every $\epsilon\in\R$ and $L>0$ it hols true that
$$
\s{N}^p_L(\epsilon)\;=\;\frac{\epsilon}{2\pi}\;.
$$
\end{lemma}
\proof
By introducing the local Fourier basis supported in $[-L,L]$
$$
\psi^L_n(x)\;:=\;\frac{\chi_{[-L,L]}(x)}{\sqrt{2L}}\;\expo{\ii\pi\frac{n}{L}x}\;,\qquad\quad n\in\Z
$$ 
one obtains that
\begin{equation}\label{eq:xoxo_=1}
\begin{aligned}
\s{N}^p_L(\epsilon)\;&=\;\frac{{\rm sgn}(\epsilon)}{2L}
\sum_{n\in\Z}\langle\psi^L_n, P_\epsilon \psi^L_n\rangle\;=\;
\frac{1}{2L}
\sum_{n\in\Z}\int_{0}^\epsilon\mu_{\psi_n^L}^{p}(\dd\epsilon')\\
&=\;\frac{1}{2L}\sum_{n\in\Z}\int_{0}^\epsilon\dd \epsilon'\;|\widehat{\psi}_n^L(\epsilon')|^2\;=\;\int_{0}^\epsilon\dd\epsilon'\; g_L(\epsilon')
\end{aligned}
\end{equation}
where 
\begin{equation}\label{eq:xoxo}
\begin{aligned}
g_L(\epsilon)\;:&=\;\frac{1}{2L}\sum_{n\in\Z}|\widehat{\psi}^L_n(\epsilon)|^2\;=\;\frac{1}{2\pi}\sum_{n\in\Z}\left(\frac{\sin(\epsilon L-\pi n)}{\epsilon L-\pi n}\right)^2\\
&=\;\frac{1}{2\pi}\left(\frac{\sin(\epsilon L)}{\pi}\right)^2\sum_{n\in\Z}{\left(\frac{\epsilon L}{\pi }-n\right)^{-2}}\;.
\end{aligned}
\end{equation}
Observe that the exchange between the sum and the integral in the last equality of \eqref{eq:xoxo_=1} is justified by the monotone convergence theorem and the computation \eqref{eq:xoxo}.
The formula $\sum_{n\in\Z}(a-n)^{-2}=(\frac{\pi}{\sin(a)})^2$ \cite{} provides
$g_L(\epsilon)=\frac{1}{2\pi}$ independently of $L$.
\qed

\medskip

The \emph{integrated density of states} (IDOS) $\s{N}^p:\R\to\R$ is defined by the limit
$$
\s{N}^p(\epsilon)\;:=\;\lim_{L\to+\infty}\s{N}^p_L(\epsilon)\;.
$$
From Lemma \ref{lemma:IDOS_p} one gets that
$$
\s{N}^p(\epsilon)\; =\;\frac{\epsilon}{2\pi}\;=\;\int_{0}^\epsilon\dd\epsilon'\; g(\epsilon')\;
$$
where the last equality emphasizes the fact that $\s{N}^p$ can be obtained by integrating the constant \emph{density of states} (DOS) $g(\epsilon):=\frac{1}{2\pi}$.

\medskip

The definition of the IDOS can be generalized allowing sequences of increasing sets less symmetric than $[-L,L]$. This essentially boils down on the invariance of $p$ under translations. 
\begin{lemma}\label{lemma:IDOS_p_02}
For every $\epsilon\in\R$ and every 
interval $\Lambda:=[a,b]\subset\R$ of finite volume $|\Lambda|=b-a$ it hols true that
$$
\s{N}^p_\Lambda(\epsilon)\;:=\;\frac{{\rm sgn}(\epsilon)}{|\Lambda|}\;{\rm Tr}\left(P_\epsilon Q_\Lambda\right)\;=\;\frac{\epsilon}{2\pi}\;
$$
where $Q_\Lambda$ is the projection
 on $\Lambda$.
\end{lemma}
\proof
Set $L:=\frac{b-a}{2}$ and $d:=-\frac{a+b}{2}$.
 Let $U_d$ be the unitary operator defined by $(U_d\psi)(x):=\psi(x-d)$. A simple calculation provides $U_d Q_\Lambda U_d^*=Q_{[-L,L]}\equiv Q_L$.
From the invariance of the trace under unitary equivalences and the fact that $P_\epsilon$ and $U_d$ commute one gets
$$
\s{N}^p_\Delta(\epsilon)\;=\;\frac{{\rm sgn}(\epsilon)}{|\Lambda|}\;{\rm Tr}\left(U_d P_\epsilon Q_\Lambda U_d^*\right)\;=\;\frac{{\rm sgn}(\epsilon)}{2L}\;{\rm Tr}\left( P_\epsilon  Q_L\right)\;=\;\s{N}^p_L(\epsilon)\;.
$$ 
The claim follows from Lemma \ref{lemma:IDOS_p}.

\begin{remark}[The DOS of the Laplacian]
{\upshape
The IDOS of the momentum $p$ and of the Laplacian $p^2$ are easily related  by observing that
$\chi_{[0,\epsilon]}(x^2)=\chi_{[-\sqrt{\epsilon},\sqrt{\epsilon}]}(x)$. From this  relation one deduces
$$
\s{N}^{p^2}(\epsilon)\;=\;\s{N}^{p}\big(\sqrt{\epsilon}\big)-\s{N}^{p}\big(-\sqrt{\epsilon}\big)\;=\;2\s{N}^{p}\big(\sqrt{\epsilon}\big)\;=\;\frac{\sqrt{\epsilon}}{\pi}\;,\quad \epsilon\geqslant 0\;.
$$ 
The last equality allows  to recover the well-known formula for the DOS of the Laplacian which is given by $g^{(2)}(\epsilon):=\frac{1}{2\pi}\frac{1}{\sqrt{\epsilon}}$.}
  \hfill $\blacktriangleleft$
\end{remark}

%--------------------%
%--------------------%
\section{Technical tools}
\label{app:tech_tool}

%------%
\subsection{Some principal value integrals}
\label{sec:comp_ker_B}
The central argument of this appendix is the study of the  following \emph{principal value} integral
$$
\s{P}\int_\R\dd u\; f(u)\;:=\;\lim_{\substack{R\to+\infty\\
r\to0^+
}} \int_{\n{I}_{R,r}}\dd u\; f(u) 
$$ 
where $\n{I}_{R,r}:=[-R,-r]\cup[+R,+r]$ for all $R>r>0$.
\begin{lemma}\label{lemma:main_cauch}
Let 
$$
G_s^\pm(u)\;:=\; \frac{\expo{\ii s\lp u\pm\frac{1}{u}\rp}}{u}\;,\qquad s\in\R.
$$
Then the principal value of $G_s^\pm$ is given by
\begin{equation}\label{eq:form_resid}
\s{P}\int_\R\dd u\;G_s^\pm(u)\;=\;\ii (1\pm 1)\pi\; {\rm sgn}(s)\;J_0\left(2|s|\right)
\end{equation}
where  $J_0$ is the  0-th Bessel function of the first kind.
\end{lemma}
\proof
For the trivial case $s=0$ one has that $G_0^\pm(u)=u^{-1}$ and  
$$
\int_{\n{I}_{R,r}}\frac{\dd u}{u}\; =\;0\;,\qquad \forall\; R>r>0
$$
since the function $u^{-1}$ is odd and the integration domain $\n{I}_{R,r}$ is symmetric with respect the origin. It follows that the principal value of $G_0^\pm$ is identically zero according to \eqref{eq:form_resid}. For $s\neq 0$ we one has the symmetry
$$
G_{-|s|}^\pm(u)\;=\; -\;G_{|s|}^\pm(-u)
$$
which provides
\begin{equation}\label{eq:first_sym}
\s{P}\int_\R\dd u\;G_{-|s|}^\pm(u)\;=\;\s{P}\int_\R\dd (-u)\;G_{|s|}^\pm(-u)\;=\;-\s{P}\int_\R\dd u\;G_{|s|}^\pm(u)\;.
\end{equation}
The relation \eqref{eq:first_sym} guarantees that we can  focus only on the case $s>0$.
In this case the computation of the principal value of $G_s^\pm$ requires the 
Cauchy's residue theorem. The function $G_s^\pm$ has a holomorphic extension to every bounded open subset of $\C\setminus\{0\}$ and has a singularity in $0$. 
Let us start by computing the residue of $G_s^\pm$. From the formula of the generating function for Bessel functions \cite[eq. 8.511 (1)]{gradshteyn-ryzhik-07} one obtains the Laurent serie
$$
\begin{aligned}
G_s^{-}(u)\;&=\;\sum_{n\in\Z}J_n(\ii2s)\;u^{n-1}\;=\;\sum_{n\in\Z}\ii^n I_n(2s)\;u^{n-1}\;
\end{aligned}
$$
where the $J_n$ are the Bessel function of the first kind and the $I_n(z):=(-\ii)^nJ_n(\ii z)$ 
are the modified Bessel functions of the first kind.
The Laurent serie for $G_s^{+}$ can be derived from the relation 
$$
G_s^{+}(u)\;=\;\ii G_{-\ii s}^{-}(\ii u)
$$
and provides
$$
\begin{aligned}
G_s^{+}(u)\;&=\;\sum_{n\in\Z}  \ii^n\;J_n(2s)\;u^{n-1}\;.
\end{aligned}
$$
By definition, the residue of $G_s^\pm$ is the coefficient of its  
Laurent series for $n={-1}$. This provides
$$
{\rm Res}_{u=0}(G_s^-)\;=\;I_0(2s)\;,\qquad {\rm Res}_{u=0}(G_s^+)\;=\;J_0(2s)\;.
$$
From the Cauchy's residue theorem one gets
$$
\ii 2\pi{\rm Res}_{u=0}(G_s^\pm)\;=\;\oint_{\Gamma_{R,r}}\dd z\; G_s^\pm(z)\;=\;\left(\int_{\n{I}_{R,r}}+\int_{C_R^+}+\int_{C_r^-}\right)\dd z\; G_s^\pm(z)
$$
where $\Gamma_{R,r}$ is a positively (counterclockwise) oriented simple closed curve
composed by the union of the domain $\n{I}_{R,r}$ on the real line, the semicircle $C_r^-:=\{r\expo{\ii\theta}\; |\; 
\theta\in[-\pi,0]\}$ in the lower half-plane and the semicircle 
${C}_R^+:=\{R\expo{\ii\theta}\; |\; 
\theta\in[0,\pi]\}$ in the upper half-plane.
An explicit computation provides
$$
\int_{\s{C}_R^+}\dd z\; G_s^\pm(z)\;=\;\ii \int^{+\pi}_0\dd\theta\;
\expo{\ii s \lp R\pm R^{-1}\rp\cos\theta}
 \expo{- s\lp R\mp R^{-1}\rp\sin\theta}\;.
$$
and consequently one has the following estimate
$$
\left|\int_{\s{C}_R^+}\dd z\; G_s^\pm(z)\right|\;\leqslant\; \int^{+\pi}_0\dd\theta\;
 \expo{- s\lp R\mp R^{-1}\rp\sin\theta}\;.
$$
Since $\expo{- s\lp R\mp R^{-1}\rp\sin\theta}\to 0$ when $R\to+\infty$ for all $\theta\in(0,\pi)$, it follows from the Lebesgue's dominated convergence theorem that
\begin{equation}\label{eq:cauch_01}
\lim_{R\to+\infty}\int_{\s{C}_R^+}\dd z\; G_s^\pm(z)\;=\;0\;.
\end{equation}
A similar computation for the integral along $\s{C}_r^-$ provides
$$
\int_{\s{C}_r^-}\dd z\; G_s^\pm(z)\;=\;\ii \int_{-\pi}^0\dd\theta\;
\expo{\ii s \lp r\pm r^{-1}\rp\cos\theta}
 \expo{- s\lp r\mp r^{-1}\rp\sin\theta}\;.
$$
After the change of coordinate $\theta\mapsto -\theta$ one gets
$$
\left|\int_{\s{C}_r^-}\dd z\; G_s^\pm(z)\right|\;\leqslant\; \int^{+\pi}_0\dd\theta\;
 \expo{ \mp s\lp r^{-1}\pm r\rp\sin\theta}\;.
$$
The latest inequality along with the Lebesgue's dominated convergence theorem provides 
\begin{equation}\label{eq:cauch_02}
\lim_{r\to0^+}\int_{\s{C}_r^-}\dd z\; G_s^+(z)\;=\;0
\end{equation}
but we didn't get a similar result for $G_s^-(z)$. Putting together \eqref{eq:cauch_01}, \eqref{eq:cauch_02} and the formula of the residue theorem one gets
\begin{equation}\label{eq:cauch_0333}
\s{P}\int_\R\dd u\;G_s^+(u)\;=\;\ii2\pi J_0(2s)\;,\qquad s>0\;
\end{equation}
Finally,from both estimates and the residue the following uniform bound for $r<1<R$ is obtained
\begin{equation*}
	\left|\int_{\n{I}_{R,r}}\dd z\; G_s^+(z)\right|\leqslant 4\pi
\end{equation*}
For the case $s<0$ the relation \eqref{eq:first_sym} immediately provides
\begin{equation}\label{eq:cauch_04}
\s{P}\int_\R\dd u\;G_s^+(u)\;=\;-\ii2\pi J_0(2|s|)\;,\qquad s<0\;.
\end{equation}
Equations \eqref{eq:cauch_0333} and \eqref{eq:cauch_04} together,  provide the proof of the formula \eqref{eq:form_resid} for $G_s^+$ (which automatically includes also the case $s = 0$ discussed at the beginning). The case of  $G_s^-$ can be managed by the following application of the Cauchy's residue theorem
$$
0\;=\;\oint_{\Sigma_{R,r}}\dd z\; G_s^-(z)\;=\;\left(\int_{\n{I}_{R,r}}+\int_{C_R^+}-\int_{C_r^+}\right)\dd z\; G_s^-(z)
$$
where $\Sigma_{R,r}$ is a positively (counterclockwise) oriented simple closed curve
composed by the union of the domain $\n{I}_{R,r}$ on the real line the semicircles 
${C}_R^+:=\{R\expo{\ii\theta}\; |\; 
\theta\in[0,\pi]\}$ and ${C}_r^+:=\{r\expo{\ii\theta}\; |\; 
\theta\in[0,\pi]\}$
both
in the upper half-plane. The zero on the right-hand side is justified by the fact that $\Sigma_{R,r}$ does not enclose the singularity of $G_s^-(z)$ and the negative sign on the last integral is due to the fact that the  semicircle ${C}_r^+$ ha the opposite orientation with respect to ${C}_R^+$. Equation \eqref{eq:cauch_01} takes care of the integral over ${C}_R^+$. The integral over ${C}_r^+$ can be controlled by observing that
$$
\left|\int_{\s{C}_r^+}\dd z\; G_s^\pm(z)\right|\;\leqslant\; \int_{0}^\pi\dd\theta\;
 \expo{- s\lp r+ r^{-1}\rp\sin\theta}\;.
$$
and, in turn
\begin{equation}\label{eq:cauch_022}
\lim_{r\to0^+}\int_{\s{C}_r^+}\dd z\; G_s^-(z)\;=\;0\;,\qquad s>0
\end{equation}
as a consequence of the
 Lebesgue's dominated convergence theorem.
Putting together \eqref{eq:cauch_01}, \eqref{eq:cauch_022}  in the Cauchy's residue formula one gets
\begin{equation}\label{eq:cauch_03}
\s{P}\int_\R\dd u\;G_s^-(u)\;=\;0\;,\qquad s>0\;. 
\end{equation}
Similarly to case (a), an analogous bound can also be obtained. Combining both results we have
\begin{equation}\label{eq:precota}
\left|\int_{\n{I}_{R,r}}\dd z\; G_s^\pm(z)\right|\leqslant 4\pi.
\end{equation}
These same results also hold also for  $s < 0$ in view of the relation \ref{eq:first_sym}.\qed

\begin{corollary}\label{corol:kern_main}
The formula 
$$
\s{P}\int_\R\dd u\;\frac{\expo{\ii xu}\expo{-\ii \frac{y}{u}}}{ u}\;=\;\ii 2\pi\left(\frac{{\rm sgn}(x)-{\rm sgn}(y)}{2}\right)\;J_0\left(2\sqrt{|xy|}\right)
$$
holds true for all $(x,y)\in\R^2$.
Moreover the following uniform bound
\begin{equation}\label{eq:cotaPV}
\left|\int_{\n{I}_{R,r}}\dd u\;\frac{\expo{\ii xu}\expo{-\ii\frac{ y}{u}}}{u}\right|\leqslant 4\pi 
\end{equation}
Is valid $\forall\; x,y\in \R$.
\end{corollary}
\proof
Let us start by considering the singular situations $xy=0$. The case $x=0=y$ corresponds to 
$$
\s{P}\int_\R\frac{\dd u}{u}\;=\;0
$$
as proved at the beginning of Lemma \ref{lemma:main_cauch}. The case $y=0$ is proportional to the (well known) Fourier transform of the function $u^{-1}$ and provides
$$
\s{P}\int_\R\dd u\;\frac{\expo{\ii xu}}{ u}\;=\; -\sqrt{2\pi}\;\bb{F}\left(\frac{1}{u}\right)\;=\;\ii\pi\;{\rm sgn}(x)\;.
$$
The case $x=0$ can be  treated with the change of variables $u\mapsto -v^{-1}$ which provides
$$
\s{P}\int_\R\dd u\;\frac{\expo{-\ii \frac{y}{u}}}{ u}\;=\;-\s{P}\int_\R\dd v\;\frac{\expo{\ii yv}}{ v}\;=\;-\ii\pi\;{\rm sgn}(y)\;.
$$
The non singular situation $xy\neq 0$ can be separated in  two different cases: (a) $xy>0$, and (b) $xy<0$.

\medskip
\noindent
{\bf Case (a).} Let  $a:=\sqrt{xy}$. Then, after  the change of variables $v:=\frac{a}{|y|}u$, one has
\begin{align*}
 \int_{\n{I}_{R,r}}\dd u\;\frac{\expo{\ii xu}\expo{-\ii\frac{ y}{u}}}{u}\;=\;\int_{{\n{I}}_{R',r'}}\dd v\;\frac{\expo{\ii\frac{x |{y}|}{a}v}\expo{-\ii{\rm sgn}(y)\frac{a}{v}}}{v}\;=\;\int_{{\n{I}}_{R',r'}}\dd v\; G_s^-(v)
\end{align*}
where $R':=a|y|^{-1}R$, $r':=a|y|^{-1}r$
and $s=a\;{\rm sgn}(y)$. Then, Lemma \ref{lemma:main_cauch} provides
$$
\s{P}\int_\R\dd u\;\frac{\expo{\ii xu}\expo{-\ii \frac{y}{u}}}{ u}\;=\;\s{P}\int_\R\dd v\;G_s^-(v)\;=\;0\;.
$$

\medskip
\noindent
{\bf Case (b).} Let  $b:=\sqrt{|xy|}$. Then, after  the change of variables $v:=\frac{b}{|y|}u$, one has
\begin{align*}
 \int_{\n{I}_{R,r}}\dd u\;\frac{\expo{\ii xu}\expo{-\ii\frac{ y}{u}}}{u}\;=\;\int_{\n{I}_{R',r'}}\dd v\;\frac{\expo{\ii\frac{x |{y}|}{b}v}\expo{-\ii{\rm sgn}(y)\frac{b}{v}}}{v}%
 \;=\;\int_{\n{I}_{R',r'}}\dd v\; G_s^+(v)
\end{align*}
where $R':=b|y|^{-1}R$, $r':=b|y|^{-1}r$
and $s=-b\;{\rm sgn}(y)$. Again 
Lemma \ref{lemma:main_cauch} provides
$$
\s{P}\int_\R\dd u\;\frac{\expo{\ii xu}\expo{-\ii \frac{y}{u}}}{ u}\;=\;\s{P}\int_\R\dd v\;G_s^+(v)\;=\;-\ii 2\pi {\rm sgn}(y)J_0(2\sqrt{|xy|})\;.
$$
The observation that $-2{\rm sgn}(y)={\rm sgn}(x)-{\rm sgn}(y)$ when $xy<0$ completes this case. The uniform bound \eqref{eq:cotaPV} is deduced directly from
\eqref{eq:precota} and the particular case $x=y=0$
\qed

%-----%
%------%
\subsection{Irregular Kelvin functions}
\label{sec:irr_kelvin}
A reference for the \emph{(irregular) Kelvin functions} is \cite[Chapter 55]{oldham-myland-spanier-09}. Here we are interested only on the 
irregular functions of $0$-th order
$$
{\rm ker}(x)\;:=\;{\rm ker}_{\nu=0}(x)\;,\qquad
{\rm kei}(x)\;:=\;{\rm kei}_{\nu=0}(x)\;.
$$
We are interested in the  behavior of these functions on the half line $\R_+:=[0,+\infty)$.
Both ${\rm ker}(x)$ and ${\rm kei}(x)$ have an exponential decay of the type $\sim\sqrt{\frac{\pi}{2x}}\expo{-\frac{x}{\sqrt{2}}}$ when $x\to+\infty$. The function 
${\rm kei}(x)$ is regular in the origin where it takes the value ${\rm kei}(0)=-\frac{\pi}{4}$. The function 
${\rm ker}(x)$ diverges at the origin as $\sim -\log(x)$. In particular one has that both the  
Kelvin functions are in $L^2(\R_+)$.
The importance of the Kelvin functions for the present work is related to the next result.

\begin{lemma}\label{lemma:spec_int_funct_appB}
Let $\bb{B}(x,y)$ the kernel \eqref{eq:int_ker_B}. Then, the following formulas hold true:
$$
\begin{aligned}
\int_\R\dd y\; \frac{\bb{B}(x,y)}{1+y^2}\;&=\;-\ii2\;{\rm sgn}(x)\;{\rm kei}\left(2\sqrt{|x|} \right)\\
\int_\R\dd y\; \frac{\bb{B}(x,y)\;y}{1+y^2}\;&=\;-\ii2\;{\rm ker}\left(2\sqrt{|x|} \right)
\\
\end{aligned}
$$
\end{lemma}
\proof
After the change of variable $s:=xy$ one gets
$$
\begin{aligned}
\s{I}_1(x)\;:=\;\int_\R\dd y\; \frac{\bb{B}(x,y)}{1+y^2}\;&=\;\ii x\int_{-\infty}^0\dd{s}\; \frac{J_0\left(2\sqrt{|s|}\right)}{x^2+s^2}\;.
\end{aligned}
$$
A second change of variable $s:=-t^2$ provides 
$$
\begin{aligned}
\s{I}_1(x)\;&=\;\ii 2x\int^{+\infty}_0\dd{t}\;t \frac{J_0\left(2t\right)}{x^2+t^4}\\
&=\;\ii2\;{\rm sgn}(x)\int^{+\infty}_0\dd\left(\frac{t}{\sqrt{|x|}}\right)\;\left(\frac{t}{\sqrt{|x|}}\right) \;\frac{J_0\left(2\sqrt{|x|} \frac{t}{\sqrt{|x|}}\right)}{\left(\frac{t}{\sqrt{|x|}}\right)^4+1}\\
&=\;-\ii2\;{\rm sgn}(x)\;{\rm kei}\left(2\sqrt{|x|} \right)
\end{aligned}
$$
where the last equality is justified by \cite[eq. 55:3:6]{oldham-myland-spanier-09}.\\ 
The second formula can be proved with similar changes of variable and one gets
$$
\begin{aligned}
\s{I}_2(x)\;:&=\;\int_\R\dd y\; \frac{\bb{B}(x,y)\; y}{1+y^2}\;=\;\ii\int_{-\infty}^0\dd{s}\; \frac{J_0\left(2\sqrt{|s|}\right) s}{x^2+s^2}\\
&=\;-\ii2\int^{+\infty}_0\dd{t}\;t^3 \frac{J_0\left(2t\right)}{x^2+t^4}\\
&=\;-\ii2\int^{+\infty}_0\dd\left(\frac{t}{\sqrt{|x|}}\right)\;\left(\frac{t}{\sqrt{|x|}}\right)^3 \;\frac{J_0\left(2\sqrt{|x|} \frac{t}{\sqrt{|x|}}\right)}{\left(\frac{t}{\sqrt{|x|}}\right)^4+1}\\
&=\;-\ii2\;{\rm ker}\left(2\sqrt{|x|} \right)
\end{aligned}
$$
where the last equality comes from \cite[eq. 55:3:5]{oldham-myland-spanier-09}.
\qed

%-----%
%------%
\subsection{Bessel equation and Hankel transform}\label{sect:hank_tras}
According to  \eqref{eq:intro_10},
the eigenvalue  equation
associated to the  one-dimensional version of the operator $T$ is 
\begin{equation}\label{eq:Bess_eq_01}
x\frac{\dd^2\psi}{\dd x^2}(x)\;+\; \frac{\dd\psi}{\dd x}(x)\;=\;-k\psi(x)\;,\qquad k\in\R\;.
\end{equation}
The change of coordinates $x(u,k):=\frac{u^2}{4k}$ produces
\begin{equation}\label{eq:Bess_eq_02}
\frac{\dd^2\phi}{\dd u^2}(u)\;+\;\frac{1}{u}\frac{\dd\phi}{\dd u}(u)\;+\;\phi(u)\;=\;0
\end{equation}
where $\phi(u):=\psi(x(u,k))$. The   \eqref{eq:Bess_eq_02} are the Bessel's equations
of order $0$-th and the solutions are the function $J_0(u)$ and $Y_0(u)$ in the standard case and $K_0(u)$ and $I_0(u)$ in the modified case. The only solution which  has no singularity is the $J_0$. 
With this information, a physical (a.k.a. non singular) solution of \eqref{eq:Bess_eq_01} in the case $k>0$ is
$$
\psi_{k>0}(x)\;:=\;\chi_{[0,+\infty)}(x)\;J_0\left(2\sqrt{|kx|}\right)\;,
$$
while in the opposite case $k<0$ is
$$
\psi_{k<0}(x)\;:=\;\chi_{(-\infty,0]}(x)\;J_0\left(2\sqrt{|kx|}\right)\;,
$$
where  $\chi_{\s{I}}$ is the characteristic function of the interval $\s{I}$. 
In the case $k=0$ the general solution of  \eqref{eq:Bess_eq_01} is $c_1\log(|x|)+c_2$, then the  physical solution can be chosen as the constant solution
$$
\psi_{k=0}(x)\;:=\;1\;.
$$
 These  solutions are not in $L^2(\R)$
 but they meet the (generalized) normalization condition
 $$
 \int_\R\dd x\; \psi_{k}(x)\psi_{k'}(x)\;=\;\delta(k-k')
 $$
 in view of \cite[6.512 (8)]{gradshteyn-ryzhik-07}.  
  Let $f\in L^1(\R)$ and  define the generalized eigenfunction expansion
$$
\begin{aligned}
\psi_f(x)\;:&=\;\int_\R\dd k\;\psi_{k}(x) f(k)\\
&=\;\chi_{(-\infty,0]}(x)\;(\bb{H}_-f)(x)\;+\;\chi_{[0,+\infty)}(x)\;(\bb{H}_+f)(x)
\end{aligned}
$$
where
$$
(\bb{H}_\pm f)(x)\;:=\;\int^{+\infty}_0\dd k\;J_0\left(2\sqrt{|kx|}\right) f(\pm k)
$$
are (a variant of) the Hankel transform of $f$ \cite[p. 3]{erdelyi-54}. For $f\in L^2(\R)$ it is possible to prove that $\psi_f\in L^2(\R)$.
In this way the Hankel transform can be used to generalize the Fourier theory for the operator
\eqref{eq:intro_11}.

\medskip

As a final remark, it is worth observing that the kernel  \eqref{eq:impl_int_ker_res} of the resolvent $(T-\alpha{\bf 1})^{-1}$ can be obtained by the expansion on the basis $\psi_k$ according to 
$$
\bb{Z}_\alpha(x,y)\;=\;\int_\R\dd k\; \frac{\psi_k(x)\psi_k(y)}{k-\alpha}\;,\qquad \alpha\in\C\setminus\R\;.
$$
The last expression can be integrated by means of the formulas \cite[eq. 6.541 (1)]{gradshteyn-ryzhik-07}.

%------------------------------------------------------------------------------------------------------------------------%
%                                                                        bibliography
%------------------------------------------------------------------------------------------------------------------------%

%------------------------------------------------------------------------------------------------------------------------%
\end{document}